\documentclass{aastex631}
\usepackage{amsmath}
\usepackage{graphicx}
\usepackage[version=4]{mhchem}
\usepackage{here}

\makeatletter

\def\acknowledgments{
\global\setbox\ackbox=\vbox\bgroup 
\vskip 5.8mm plus 1mm minus 1mm
\vskip1sp
\noindent\ignorespaces 
}

\def\endacknowledgments{
\egroup
\ifanonymous
\vskip 5.8mm plus 1mm minus 1mm
\vskip1sp
\centerline{(Acknowledgments anonymized for review)}
\else
\vbox{\unvbox\ackbox}
\fi\vskip6pt
}

\makeatother

\accepted{ApJ}

\begin{document}
\author[0009-0007-3387-9495]{Prabal Saxena}
\affiliation{NASA Goddard Space Flight Center,
8800 Greenbelt Road
Greenbelt, MD 20771, USA}

\author[0000-0002-5967-9631]{Thomas J. Fauchez}
\affiliation{NASA Goddard Space Flight Center,
8800 Greenbelt Road
Greenbelt, MD 20771, USA}
\affiliation{Integrated Space Science and Technology Institute, Department of Physics, American University, Washington DC}

\title{Modeling Volcanic Plume Heights Across Exoplanet Atmospheres: Insights from TRAPPIST-1}

\begin{abstract}

Explosive volcanic eruptions play a fundamental role in the evolution and observability of rocky exoplanets, serving as a key mechanism for injecting volatiles into planetary atmospheres and potentially modifying their climate and composition. This process may be particularly important for close-in exoplanets where tidal forcing can drive substantial internal heating, analogous to (but often exceeding) Io's volcanism. In this work, we adapt and extend a classic 1D volcanic plume model originally developed in IDL by Glaze and Baloga for Venus and Mars applications, and port it into a flexible, open Python framework suitable for exoplanet studies. The model explicitly couples vent thermodynamics, buoyant entrainment, and vertically varying static stability to predict plume rise, neutral-buoyancy height, and overshoot for a wide range of planetary and atmospheric conditions.

We first benchmark the Python implementation against the original IDL code and against analytic scaling laws to ensure adequate momentum budgets, and strict mass conservation. We then apply the model to a suite of exoplanet-relevant background states, including CO$_2$-rich atmospheres under strong irradiation and diverse surface conditions. A systematic sensitivity analysis explores how plume height depends on surface gravity, bulk atmospheric composition (and mean molecular weight), background temperature and stratification, vent overpressure, and volatile loading. We identify regions of parameter space where plumes routinely penetrate to low-pressure levels, maximizing their potential detectability in transmission or emission. These results provide a physically grounded framework for predicting whether and how volcanic emissions might be detected on rocky exoplanets, including but not limited to those experiencing strong tidal heating.

\end{abstract}



\section{Introduction} \label{sec:intro}

Volcanic outgassing has been recognized as a fundamental process shaping the evolution and observability of rocky exoplanets. Secondary atmospheres—which may dominate the atmospheric inventories of terrestrial planets in the habitable zone—are inherently linked to volcanic activity, as ongoing or episodic magmatic degassing replenishes volatiles lost to space and drives long-term atmospheric composition over billions of years \citep{gaillard2021diverse, kite2020exoplanet, ortenzi2020mantle}. The composition of outgassed volatiles depends critically on mantle redox state, with oxidized mantles likely producing H$_2$O- and CO$_2$-rich gases while reduced conditions favor CH$_4$ and CO \citep{ortenzi2020mantle}. For tidally heated exoplanets specifically, theoretical predictions of heat flow \citep{Jackson2008, Henning2009, henning2018highly, quick2020forecasting} suggest that volcanism comparable to or exceeding Io's activity should be common among close-in rocky planets, with potential for sustained magma ocean surfaces on the most extremely heated worlds. Conversely, direct observational signatures of active volcanism—including transient SO$_2$ absorption features \citep{misra2015transient, loftus2019sulfate}, volcanic aerosol hazes \citep{kaltenegger2010detecting}, and thermal anomalies from lava flows—have been proposed as tracers of both atmospheric chemistry and interior properties accessible to current and near-future telescopes. Recent JWST observations have elevated these theoretical predictions to testable hypotheses, with detections of sulfur-bearing species in hot rocky exoplanet atmospheres \citep{Tsai2023, Powell_2024, bello2025evidence, Gressier_2024, Banerjee_2024} suggesting that volcanic processes may already be observationally constrained in select cases. Yet a critical gap remains between these two research frontiers—long-term outgassing models and remote observability studies—namely, the dynamics of explosive volcanic plumes themselves. These plumes are the physical mechanism by which interior volatiles are injected into planetary atmospheres, and their explosivity determines whether volcanic material reaches the pressure levels probed by transmission and emission spectroscopy \citep{Burrows2014, Irwin2014, Molliere2017}. On Earth, explosive eruptions remain frequent and globally significant \citep{calvari2020vei}, and similar processes should be expected on tectonically or tidally active rocky exoplanets. Here we address this gap by applying a well-validated, first-principles volcanic plume model—originally developed for Solar System bodies \citep{glaze1996sensitivity, glaze1997transport, Baloga_2003, Glaze_2003, Glaze_2007}—to explore the dynamics of explosive eruptions across a systematic parameter space of exoplanet atmospheric conditions. We investigate these eruption scenarios through the fiducial cases of TRAPPIST-1d and TRAPPIST-1e, two potentially rocky planets with relatively temperate atmospheres that are prime targets for habitability assessment and ongoing JWST characterization \citep{Gillon_2017, Greene_2023, Zieba_2023, ducrot_2024}.

The TRAPPIST-1 system \citep{Gillon_2017} exemplifies the potential for sustained volcanic activity on tidally heated rocky exoplanets. Strong tidal forcing in such ultra-compact multi-planet systems can generate internal heating comparable to or exceeding that of Io \citep{Barr2018, Dobos2019, Nicholls2025}, although the partitioning between observable volcanic outgassing versus interior heat dissipation through mantle convection, lithospheric conduction, or subsurface ocean/crust viscoelasticity remains uncertain. To date, no thick atmosphere has been conclusively detected around TRAPPIST-1b or 1c \citep{Greene_2023, Zieba_2023, ducrot_2024}, but the diversity in orbital distance, surface gravity (0.6--0.8 $g_{\oplus}$ for planets d--g), and predicted tidal heating rates makes the system an exceptional laboratory for exploring how planetary parameters control volcanic plume dynamics and atmospheric injection efficiency. TRAPPIST-1d and 1e are particularly compelling targets as both lie near or within the habitable zone with surface pressures potentially in the 0.5--2 bar range if secondary atmospheres are sustained \citep{Wolf_2022}, and their intermediate tidal heating rates suggest volcanic activity is plausible without necessarily producing Io-like magma ocean conditions. Even without long-lived thick atmospheres, vigorous volcanism could generate transient, patchy SO$_2$-dominated atmospheres—analogous to Io's tenuous, spatially variable envelope \citep{Trumbo_2022}—where localized volcanic plumes observed with HST \citep{Spencer_1997} and JWST \citep{dePater_2025} sustain detectable signatures despite rapid sublimation and re-condensation cycles. By coupling a validated volcanic plume model with realistic atmospheric profiles for TRAPPIST-1d and 1e derived from GCM simulations, we can directly assess whether explosive eruptions inject material to the pressure levels (10--100 mbar) where transmission and thermal emission spectra are most sensitive, providing a physically grounded framework for interpreting current and future JWST observations of this system.

In this context, we present an updated implementation of the Glaze and Baloga volcanic plume model \citep{glaze1996sensitivity, glaze1997transport}, originally developed in IDL, which we have translated into Python and generalized for exoplanet conditions. In this paper, we focus on testing and validating the model against a set of atmospheric columns for TRAPPIST-1 d \& e and  idealized atmospheres, and exploring the sensitivity of plume-top heights to key planetary and atmospheric parameters. Our primary analytical focus is on maximum plume injection heights as the key diagnostic for assessing whether volcanic material reaches observable atmospheric levels. Although the model generates comprehensive additional outputs including mass deposition profiles, entrainment fluxes, and thermodynamic evolution, detailed examination of these secondary products is reserved for future studies. These tests across these range of parameters purposely span significant phase spaces that may include atmospheric and planetary properties that may be unrealistic, in order capture sensitivity analysis trends. However, given the diversity of exoplanets observed to date, we explore both these sensitivity and validation tests that are meant to elucidate characteristics of potential explosive volcanic eruptions across a wide range of environments as well as more targeted simulations that incorporate GCM modeling that may be more realistic for the Trappist planets.  Our tests explore variations in planetary, atmospheric and volcanic eruption properties, with the eruption properties largely based off those given in previous Glaze and Baloga volcanic plume model studies. These eruption properties are inherently based on simulations that are compared to solar system observations or are based upon experimental work done relevant to these solar system applications.  We discuss why we use these solar system relevant assumptions for different parameters in both the introduction and in the description of the relevant tests that we conducted.

For example, a key assumption is that although TRAPPIST-1e and similar exoplanets may possess CO$_2$-dominated atmospheres, we assume water remains the primary volatile in explosive volcanic plumes for several reasons. First, water is approximately 5--10 times more effective at lowering silicate melting temperatures than CO$_2$ on a per-weight-percent basis \citep{dasgupta2006melting, hirschmann2006water, medard2008effect}, making it the dominant driver of partial melting in planetary mantles even when carbon species are present. Second, at the low pressures relevant to explosive volcanism ($<$3~GPa), H$_2$O solubility in basaltic melts ($\sim$6~wt\% at 2~kbar) far exceeds that of CO$_2$ ($\sim$0.1--0.5~wt\% at similar pressures; \citealt{dixon1995experimental_a, shishkina2010solubility}), meaning water preferentially partitions into ascending magmas and dominates the volatile budget during eruption. Mantle-derived melts retain more dissolved H$_2$O during ascent than CO$_2$, which exsolves at greater depth due to lower solubility \citep{dixon1995experimental_b}. Additionally, comparative planetology studies demonstrate that atmospheric pressure exerts primary control on volcanic gas composition: at surface pressures $\lesssim$1--2 bar (characteristic of TRAPPIST-1 planets and relevant to our plume source conditions), H$_2$O-dominated degassing occurs, whereas high-pressure atmospheres ($\gg$10 bar) produce CO$_2$-rich, dry volcanic gases \citep{gaillard2014theoretical, van2019exoplanet}. Notably, even Venus, despite its CO$_2$-rich atmosphere, is expected to have water as the dominant volatile in its volcanic eruptions \citep{gaillard2014theoretical, airey2015explosive}.

Carbon-dominated volcanic plumes may occur in specific scenarios, such as highly reduced planetary mantles where carbon exists as graphite or carbides rather than CO$_2$ (e.g., Mercury-like bodies; \citealt{vander2015exotic}), or in hypothetical carbon-rich super-Earths with high C/O ratios \citep{hakim2019mineralogy, liggins2022growth}. However, stellar abundance patterns and mass-radius constraints suggest rocky exoplanets are generally composed of silicate mantles similar to Solar System terrestrial planets \citep{van2019exoplanet}, and TRAPPIST-1 planets specifically show bulk densities consistent with such compositions \citep{grimm2018nature, agol2021refining}. Modeling further suggests significant water can be sequestered in nominally anhydrous mantle minerals (olivine, pyroxene) even on planets with CO$_2$-rich atmospheres \citep{luo2024interior, dorn2021hidden, moore2020keeping}, which would be released preferentially during partial melting. In the absence of observational constraints on mantle oxidation state or bulk carbon content for TRAPPIST-1e, we adopt water-dominated volcanic plumes as the baseline scenario, consistent with Solar System terrestrial planets and the thermodynamic arguments outlined above.

For our assumptions on the morphological and energetic parameters of explosive plumes, we use the Volcanic Explosivity Index (VEI) \citep{newhall1982volcanic}, an integer scale that increases roughly logarithmically with erupted tephra volume and plume height. Volcanic eruptions are inherently complex and diverse, controlled by factors including magma composition, conduit geometry, gas content, and crystallinity \citep{cassidy2018controls}, many of which are difficult to observe directly due to safety and access limitations. Individual eruptions within a single VEI classification exhibit substantial variability in vent conditions and plume dynamics \citep{koyaguchi2010effects}, and eruptive periods often consist of multiple events with varying styles. For historical eruptions, reconstructing event parameters from geological deposits remains challenging \citep{kiyosugi2015many}, making precise characterization uncertain. Despite this inherent variability, the VEI scale provides a pragmatic framework for estimating eruption magnitude, offering representative ranges of eruption and plume properties well-suited to parameter-space studies where detailed knowledge of both plume source conditions and atmospheric structure remains limited.

Table~\ref{tab:volcanic_params} summarizes the eruption parameters we adopt for VEI~2--8, including representative vent radii, exit velocities, plume temperatures, durations, and Earth analogs. On Earth, large eruptions are empirically rarer as VEI increases: VEI~2--3 events occur frequently, whereas VEI~6--7 eruptions such as Pinatubo or Tambora are separated by centuries to millennia, and VEI~8 ``super-eruptions'' are thought to be extremely rare \citep[e.g.,][]{stothers1984great,chesner1991eruptive,rampino1993climate,kandlbauer2014new,calvari2020vei,basuki20232021}. We focus on explosive, Plinian-style eruptions that inject columns of debris and gas high into the atmosphere, examining primarily VEI~4 and VEI~6 events to balance eruption intensity against plausible occurrence frequency. On Earth, VEI~4 and VEI~6 eruptions occur on roughly yearly and centennial timescales, respectively, though this cadence is subject to uncertainty due to potential underrecording in the geological record \citep{kiyosugi2015many}. As noted above, tidally forced exoplanetary systems can exhibit internal heat fluxes orders of magnitude larger than Earth's, potentially altering both eruption frequency and intensity relative to terrestrial baselines. Our VEI~4 and VEI~6 classifications therefore serve primarily to define representative source parameters (vent radius, exit velocity, temperature, volatile content -- set at 3\% by total mass for nominal water dominated cases based on \citet{glaze1996sensitivity, glaze1997transport}) rather than to predict actual eruption rates on exoplanets. These fiducial parameters are then systematically varied in our sensitivity analyses to explore the full phase space of plausible explosive volcanic conditions. We emphasize that substantial variability exists within individual VEI categories \citep{cassidy2018controls}, and our adopted values (Table~\ref{tab:volcanic_params}) represent central estimates derived from well-characterized terrestrial eruptions given in the example volcanoes column.

\begin{deluxetable}{cccccccc}
\tablecaption{Volcanic Eruption Parameters Used in this study for Different VEI Classifications \label{tab:volcanic_params}}
\tablewidth{0pt}
\tablehead{
\colhead{VEI} & 
\colhead{Ejected} & 
\colhead{Required} & 
\colhead{Vent} & 
\colhead{Plume} & 
\colhead{Plume} & 
\colhead{Duration} & 
\colhead{Example} \\
\colhead{} & 
\colhead{Tephra} & 
\colhead{Volume} & 
\colhead{Size} & 
\colhead{Velocity} & 
\colhead{Temp.} & 
\colhead{(hrs)} & 
\colhead{Volcanoes} \\
\colhead{} & 
\colhead{(m$^3$)} & 
\colhead{(m$^3$)} & 
\colhead{(m)} & 
\colhead{(m s$^{-1}$)} & 
\colhead{(K)} & 
\colhead{} & 
\colhead{}
}
\startdata
2 & $4 \times 10^6$ & $10^6$--$10^7$ & 30 & 100 & 900 & 1 & Etna\tablenotemark{a} \\
4 & $4.2 \times 10^8$ & $10^8$--$10^9$ & 100 & 250 & 900 & 6 & Semeru\tablenotemark{b} \\
5 & $4.6 \times 10^9$ & $10^9$--$10^{10}$ & 200 & 300 & 1000 & 15 & Mt. St. Helens\tablenotemark{c} \\
6 & $4.2 \times 10^{10}$ & $10^{10}$--$10^{11}$ & 500 & 400 & 1000 & 18 & Pinatubo\tablenotemark{d} \\
7 & $3.3 \times 10^{11}$ & $10^{11}$--$10^{12}$ & 1000 & 600 & 1100 & 24 & Tambora\tablenotemark{e} \\
8 & $3.1 \times 10^{12}$ & $>10^{12}$ & 2000 & 700 & 1100 & 48 & Toba\tablenotemark{f} \\
\enddata
\tablenotetext{a}{\citep{calvari2020vei, de2023assessment, calvari2024reawakening,nies2025reactive}}
\tablenotetext{b}{\citep{GVP_Semeru, basuki20232021, suhendro2025magma}}
\tablenotetext{c}{\citep{kieffer1981blast, scandone1985magma, fruchter1980mount}}
\tablenotetext{d}{\citep{koyaguchi1993origin, holasek1996satellite}}
\tablenotetext{e}{\citep{stothers1984great, kandlbauer2014new, foden1986petrology}}
\tablenotetext{f}{\citep{chesner1991eruptive, rampino1993climate, koulakov2016feeder}}
\end{deluxetable}

The remainder of this paper describes our implementation of the Glaze-Baloga model and its application to exoplanet-relevant parameter space. Section~2 summarizes the core plume physics and details our modifications, including enhanced atmospheric input capabilities, parametric VEI implementation, and improved diagnostic outputs. Section~3 presents validation tests, systematic parameter sweeps exploring sensitivity to atmospheric and eruption properties, and targeted simulations examining key atmospheric states for TRAPPIST-1d and 1e. Section~4 discusses implications for exoplanet observability and identifies priorities for future work, including coupling to global circulation models and expanded multi-planet comparative studies.





\section{Glaze-Baloga Volcanic Plume Model Implementation and Enhancements}

The volcanic plume model employed in this study builds upon the foundational four-component plume model \citep{glaze1996sensitivity, glaze1997transport}, originally developed in IDL for modeling buoyant rise of volcanic columns comprised of ash, dry air, water vapor, and liquid water. The model self-consistently solves conservation equations of mass, momentum and energy for these four components as the plume propagates upward through a stratified atmosphere. The core physics involves solving a system of five coupled differential equations describing the thermal dynamics of buoyant eruption columns, where the ambient atmosphere may have different composition than the magmatic gas. The original program takes as input initial values for gas content, plume radius, bulk column velocity, and column temperature, then solves differential equations for plume density, radius, velocity and temperature as functions of altitude using Runge-Kutta methods.

Our implementation represents an enhancement of the original framework, converted from IDL to Python with significant modifications to improve applicability to diverse exoplanetary atmospheres and eruption scenarios. The Glaze-Baloga volcanic plume model code is publicly available 
\citep{saxena2026glazecode} and can be run interactively via Binder. The adapted model includes enhanced neutral buoyancy height detection, flexible atmospheric input capabilities, systematic VEI parameterization, comprehensive mass and momentum conservation tracking, and expanded diagnostic outputs. The original model used a simple threshold-based approach for detecting neutral buoyancy height (NBH), where NBH was identified when the absolute difference between ambient and plume densities fell below 0.0001 kg/m³, often missing NBH crossings due to insufficient precision. We implemented robust sign-change detection with linear interpolation where the neutral buoyancy height is calculated as $\text{NBH} = z_{prev} + \alpha (z_{curr} - z_{prev})$, with $\alpha = \frac{\rho_{diff,prev}}{\rho_{diff,prev} - \rho_{diff,curr}}$ representing the interpolation factor, $z_{prev}$ and $z_{curr}$ the previous and current altitude steps, and $\rho_{diff} = \rho_{ambient} - \rho_{plume}$ the density difference between ambient atmosphere and volcanic plume. This method accurately captures the exact altitude where plumes transition from buoyant to negatively buoyant.

Atmospheric input capabilities were substantially enhanced to accept arbitrary pressure-temperature-humidity profiles from General Circulation Model outputs rather than being limited to predefined atmospheric models. The atmospheric interpolation function parses flexible file formats with variable headers, converts specific humidity to water vapor mixing ratio using $w_a = q/(1-q) + w_{liquid} + w_{ice}$ where $w_a$ is the water vapor mixing ratio, $q$ is the specific humidity, $w_{liquid}$ is the liquid water mixing ratio, and $w_{ice}$ is the ice water mixing ratio, and performs linear interpolation of atmospheric properties at arbitrary altitudes. Support for diverse atmospheric compositions was added, including CO$_2$-dominated, N$_2$-dominated, and mixed atmospheres relevant to exoplanetary studies, with appropriate gas constants and specific heat capacities automatically selected based on specified atmospheric composition.

A systematic approach to volcanic eruption intensity was implemented through parametric VEI classification using values given in Table~\ref{tab:volcanic_params}, providing consistent source parameters for comparative studies while allowing systematic variation during sensitivity analyses. The model includes mass conservation tracking throughout simulations, monitoring three conservation principles: solid particle mass flux (which must remain constant), magmatic water mass evolution that accounts for phase changes and entrainment, and total mass conservation including all atmospheric entrainment sources. At each integration step, entrained dry air and water vapor masses are calculated using the same derivatives employed by the Runge-Kutta integrator, ensuring exact consistency between the physics solver and conservation diagnostics. The code reports conservation errors and mass multiplication factors, providing quantitative measures of plume growth efficiency through atmospheric entrainment at each vertical step.

Momentum tracking capabilities enable analysis of how initial kinetic energy converts to gravitational potential energy during plume rise while accounting for momentum losses due to entrainment drag and density evolution. Enhanced output capabilities include detailed tabular profiles of all state variables, multi-panel diagnostic plots showing density evolution and buoyancy profiles, and comprehensive visualization of plume geometry and thermodynamic evolution. The model outputs detailed vertical profiles including density, temperature, velocity, and radius evolution, enabling analysis of plume structure and dynamics beyond simple height determination.

Our primary analytical focus is on maximum plume injection heights as the key diagnostic for assessing whether volcanic material reaches observable atmospheric levels. Although the model generates comprehensive additional outputs including mass deposition profiles, entrainment fluxes, and thermodynamic evolution, detailed examination of these secondary products is reserved for future studies. The enhanced model maintains the core physics of the original Glaze-Baloga formulation while providing improved accuracy, usability, and applicability to diverse planetary conditions and eruption scenarios relevant to exoplanet atmospheric characterization.

\section{Sensitivity of Plume Height to Environmental and Source Parameters}
\label{sec:results}

This section presents modeling results from the Glaze-Baloga plume model applied to both parameterized hypothetical atmospheres and fiducial TRAPPIST-1e and TRAPPIST-1d atmospheric profiles derived from GCM simulations. Understanding the final height of explosive volcanic plumes is critical for assessing both the atmospheric impact of eruptions and their observational signatures in transmission or emission spectra. However, plume dynamics are complex and depend on numerous atmospheric and eruption properties that interact in non-linear ways. We systematically explore how key parameters---including atmospheric pressure, temperature stratification, gravity, molecular weight, and eruption characteristics---control injection heights both individually and in combination, revealing fundamental physical regimes that govern volcanic-atmospheric coupling on exoplanets.

\subsection{Isothermal-Ambient Sensitivity of Plume Rise Height}
\label{subsec:isothermal}

We first test the sensitivity of the maximum plume rise height, $H$, to the ambient \emph{isothermal} temperature while holding the pressure profile fixed. This isolates the competition between (i) reduced source buoyancy as the ambient temperature $T_a$ increases (for fixed vent temperature $T_p$), and (ii) weakened static stability $N$ in a warmer isothermal column. For the two eruption strengths considered here, we fixed the source temperatures at $T_p=900\,\mathrm{K}$ (VEI--4) and $T_p=1000\,\mathrm{K}$ (VEI--6).

The classical Morton--Taylor--Turner framework of turbulent-plume theory \citep{morton1956turbulent} gives the first-order scaling
\begin{equation}
H \;\propto\; F_0^{1/4}\,N^{-3/4},
\label{eq:H_scaling}
\end{equation}
where $F_0$ is the source buoyancy flux and $N$ is the Brunt--V\"ais\"al\"a frequency (a measure of atmospheric static stability).

\begin{figure}
  \centering
  \includegraphics[width=0.78\textwidth]{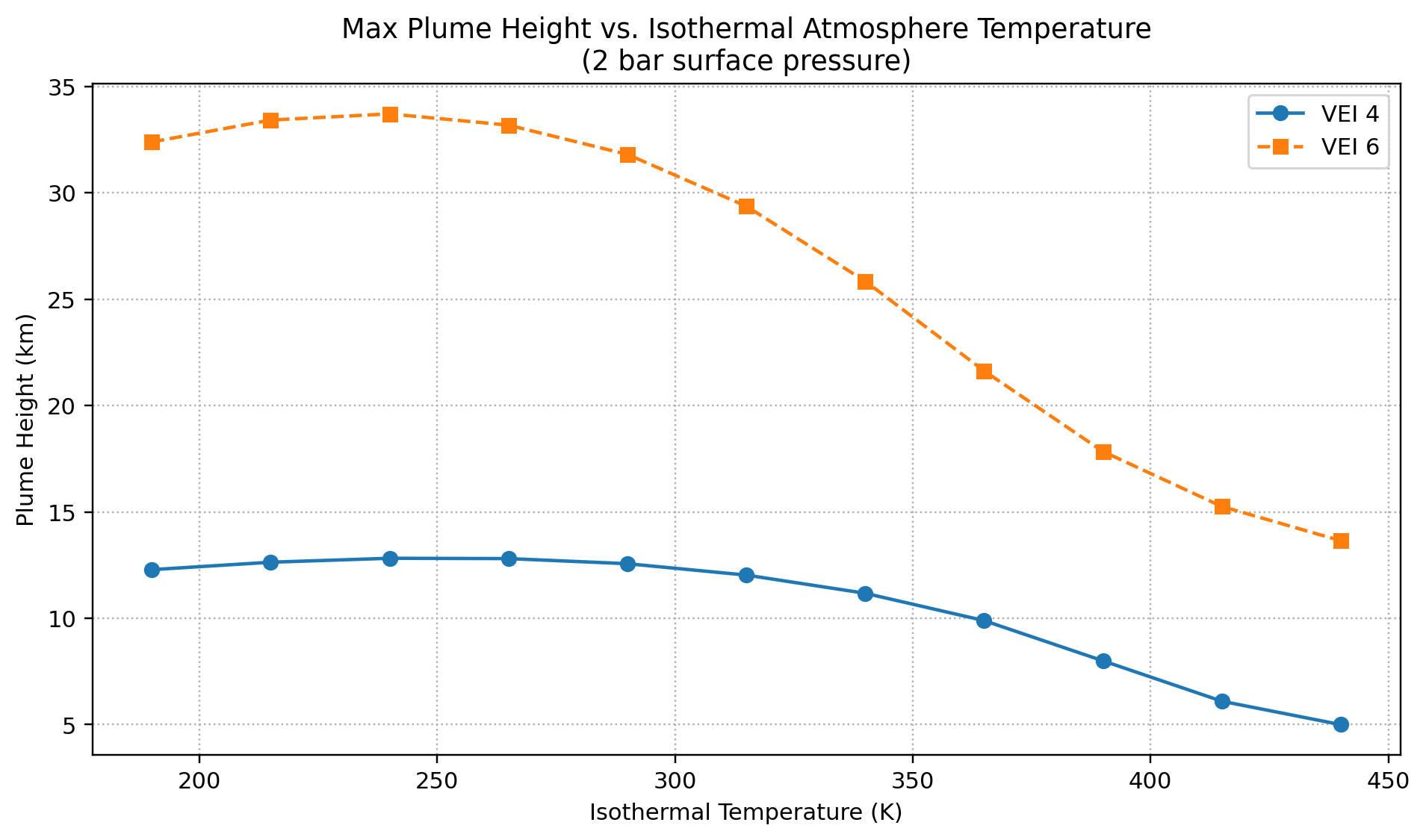}
  \caption{Plume height $H$ dependency on temperature for isothermal profiles at VEI--4 and VEI--6.}
  \label{Fig:isothermals}
\end{figure}

In our experiments, the pressure profile (and therefore the vertical stratification) is held fixed across the different isothermal atmospheres, so $N$ remains nearly unchanged. As a result, variations in plume height are primarily driven by changes in $F_0$, which depend both on the density contrast between the hot eruptive mixture and the ambient atmosphere, and on the density of the ambient fluid being entrained. Therefore, with identical source pressure and composition across cases, the buoyancy flux may be written to leading order as
\begin{equation}
F_0 \;\approx\; g\,Q_0\left(\frac{\Delta \rho}{\rho_a}\right)
\;\approx\; g\,Q_0\left(\frac{\Delta T}{T_a} - \frac{\Delta\mu}{\mu_a}\right),
\label{eq:F0}
\end{equation}
where $Q_0=\pi r_0^2 w_0$ is the source volume flux, $g$ is gravitational acceleration, $\rho_a$ is ambient density, $\Delta T$ and $\Delta\mu$ denote the source--ambient temperature and molecular–weight contrasts (here $\Delta\mu\simeq 0$). For an ideal-gas \emph{isothermal} atmosphere,
\begin{equation}
N^2 \;=\; \frac{g^2}{c_p\,T_a}\;\;\Rightarrow\;\; N \propto T_a^{-1/2}.
\label{eq:N_iso}
\end{equation}
where $c_p$ is the specific heat capacity at constant pressure. Combining Eqs.~(\ref{eq:H_scaling})--(\ref{eq:N_iso}) yields
\begin{equation}
H \;\propto\; \big(\Delta T/T_a\big)^{1/4}\; T_a^{3/8}.
\label{eq:H_combined}
\end{equation}
With $\Delta T = T_p - T_a$ and fixed source temperature $T_p$, this expression can be rewritten as
\begin{equation}
H(T_a) \;\propto\; (T_p - T_a)^{1/4}\,T_a^{1/8},
\end{equation}
which explicitly shows the competition between the decreasing buoyancy contrast $(T_p - T_a)$ and the weakly stabilizing increase of $T_a$. As a function of $T_a$, $H$ therefore does not vary monotonically: it exhibits a broad maximum when $T_a/T_p \simeq 1/3$, and decreases both toward very cold and very warm ambient states. For plume temperatures $T_p \sim 700$--$750~\mathrm{K}$, this simple scaling predicts a peak near $T_a \sim 230$--$250~\mathrm{K}$, in good agreement with the numerically simulated maximum around $T_a = 240~\mathrm{K}$ that we can notice in Fig.~\ref{Fig:isothermals} for VEI--4 and VEI--6. The dependence of the modeled plume height on the isothermal background temperature is therefore distinctly non-monotonic. 

When the background atmosphere is very cold (e.g., $T = 190~\mathrm{K}$), it is relatively dense at a given pressure. The large density contrast between the hot eruptive mixture and the cold ambient promotes strong initial buoyancy. However, entrainment of this cold, heavy air into the plume rapidly increases the plume bulk density and cools it toward the environmental temperature. This efficient dilution causes the plume to reach neutral buoyancy at a somewhat lower altitude than one might expect from the initial contrast alone. As the ambient temperature is increased toward $T \approx 240~\mathrm{K}$, the atmosphere becomes less dense, reducing the rate at which entrainment loads the plume with dense ambient gas, while the buoyancy contrast with the eruptive mixture remains sufficient to drive vigorous ascent. In this intermediate regime, the balance between sustained buoyancy and less penalizing entrainment is optimal, and the plume attains its greatest height.

For warmer isothermal backgrounds ($T \gtrsim 290~\mathrm{K}$), the ambient gas is increasingly light, and the density contrast between the plume and the environment is reduced. Even though less dense air is entrained, the diminished buoyancy contrast lowers the effective buoyancy flux $F_0$, and the plume can no longer accelerate to the same altitudes. Consequently, the neutral buoyancy level moves downward and the maximum plume height decreases with increasing temperature. The resulting plume-height curve therefore exhibits a broad maximum around $T \approx 240~\mathrm{K}$, rather than a monotonic increase or decrease with atmospheric temperature, reflecting the transition between an entrainment-limited regime at low temperatures and a buoyancy-limited regime at high temperatures under a fixed $2~\mathrm{bar}$ pressure structure.

We emphasize that the numerical value of this maximum is not universal: in our experiments it is conditioned by the assumed vent temperature, bulk atmospheric composition (and hence mean molecular weight), gravity, and source geometry and injection speed, all of which are held fixed across the temperature ensemble, so that different parameter choices would shift the optimal $T_a$ while preserving the same qualitative non-monotonic behavior.

\subsection{Stratification, Stability and Overshoot}\label{subsec:N}

In the previous subsection we examined plume behavior in a set of idealized isothermal atmospheres, where the static stability was deliberately simplified. While those experiments clarified how ambient temperature alone affects plume height, they do not capture the impact of realistic vertical variations in \(N\). Here we repeat the analysis using a dynamically consistent column generated by the ExoCAM GCM \citep{Wolf_2022} for TRAPPIST-1e assuming a 2~bar CO$_2$ atmosphere over a fully ocean (50~m deep slab) covered surface, focusing on how the diagnosed stability profile and stratification regimes control neutral buoyancy, overshoot, and the final plume-top height. 

\subsubsection{The ExoCAM GCM}

The ExoCAM simulations were run with a horizontal grid resolution of $4^\circ$ latitude by $5^\circ$ longitude and 51 vertical layers, up to a model top at 1~Pa. The radiative transfer core uses two-stream solver across 68 spectral bins combining both the shortwave (i.e. stellar) and longwave (i.e. planetary) energy streams. We selected a 2-bar CO$_2$ atmosphere as our fiducial case because this pressure is deliberately non-Earth-like while remaining in a moderate terrestrial regime, avoiding both the comparatively thin/near-Earth end-member and the highly optically thick high-pressure end-member. From a dynamical–radiative standpoint, 2~bar provides a robust separation between a convecting lower atmosphere and a more radiatively controlled, stably stratified upper atmosphere, reducing sensitivity to short-timescale variability and helping maintain a consistent cold-trap structure. 

We then selected a single atmospheric column with horizontal coordinates corresponding to TRAPPIST-1e west terminator at 45° North, a location predicted to be a surface hot spot from tidal heating (Renaud et al. 2026, in prep) with the TidalPy code (\url{https://github.com/jrenaud90/TidalPy}, \cite{Renaud_2025}).

\begin{figure}
  \centering
  \includegraphics[width=0.5\linewidth]{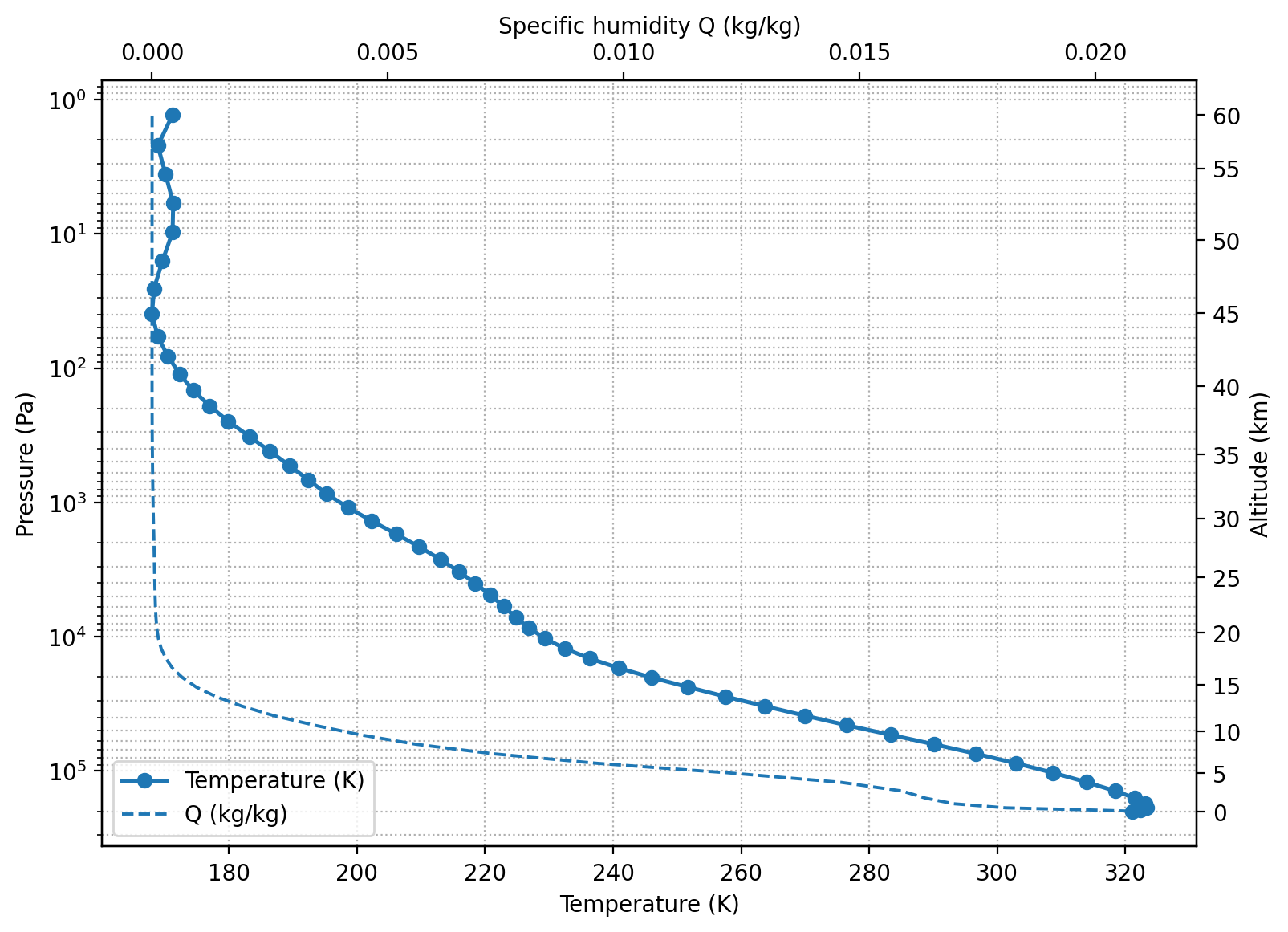}
  \includegraphics[width=0.5\textwidth]{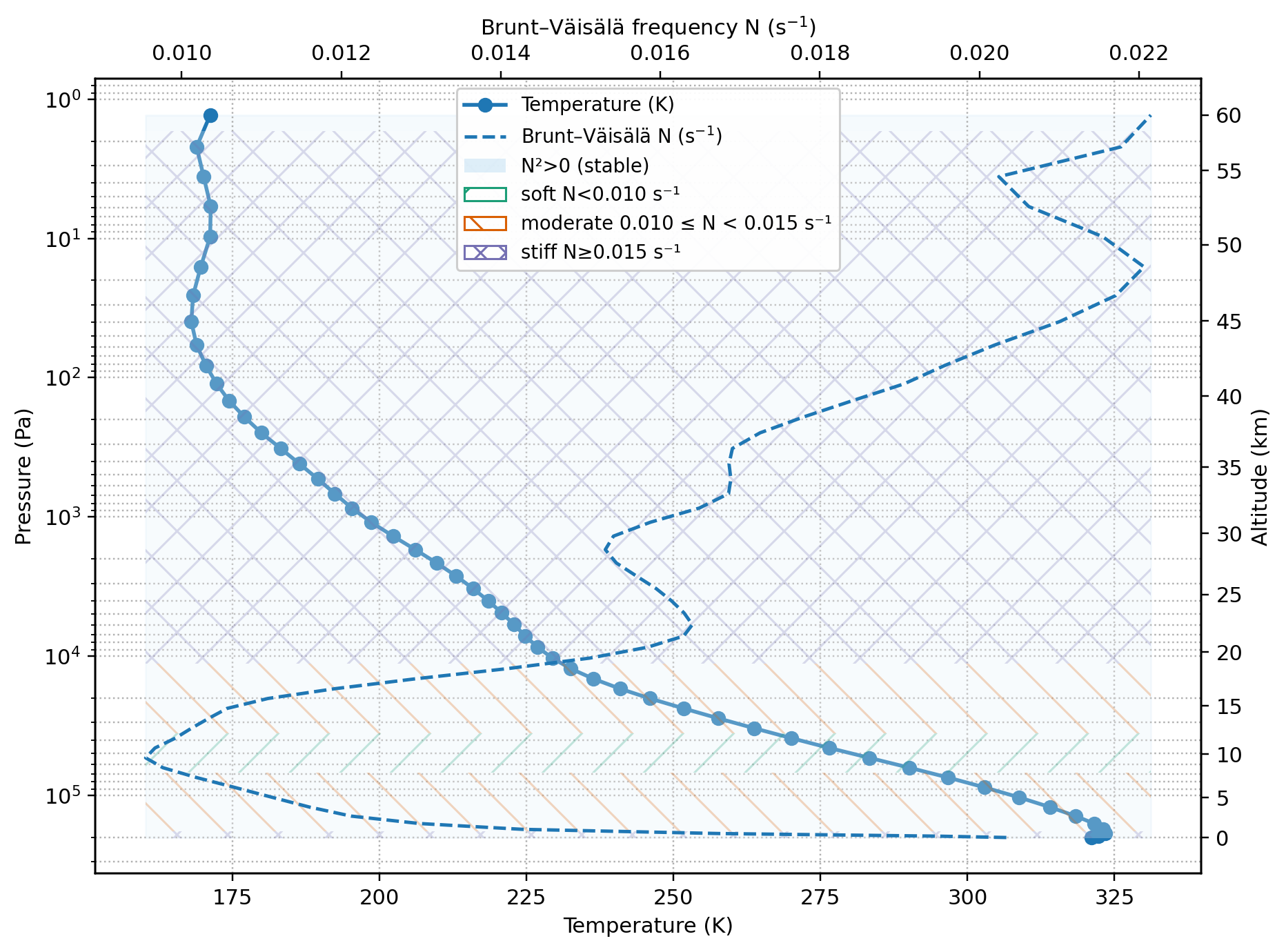}
  \caption{\textbf{Top Panel:} Background 2~bar CO$_2$ column extracted from our ExoCAM simulation of TRAPPIST-1e used in the temperature-offset experiments.
  Bottom x-axis: temperature $T$ (K). Left y-axis: pressure (Pa; logarithmic, inverted). Top x-axis: specific humidity $Q$ (kg\,kg$^{-1}$). Right y-axis: altitude (km), placed at the pressures that correspond to those altitudes so ticks align with the pressure grid.
  \textbf{Bottom Panel:} Ambient structure used in the plume run.
  Temperature vs.\ pressure (points; left abscissa/left ordinate), with pressure mapped to altitude on the right ordinate. The top abscissa gives the diagnosed Brunt--V\"ais\"al\"a frequency \(N\). Background fill encodes the sign of \(N^2\) (stable/neutral/unstable), and diagonal hatching partitions the stable column into \emph{soft} (\(N<0.010~\mathrm{s^{-1}}\), green slashes), \emph{moderate} (\(0.010\le N<0.015~\mathrm{s^{-1}}\), orange back-slashes), and \emph{stiff} (\(N\ge 0.015~\mathrm{s^{-1}}\), violet cross-hatch) regimes. The present column is everywhere stable, with a broad mid-column minimum in \(N\) and stronger stability aloft.}
  \label{fig:tpn}
\end{figure}

In Fig.~\ref{fig:tpn}, the left axis gives pressure (Pa, log, decreasing upward) versus temperature; the right axis gives the corresponding geometric altitude (km), mapped from the same pressure grid. On the upper panel, the top axis shows the specific humidity $Q$ (kg\,kg$^{-1}$), as water vapor naturally arise in the atmosphere from surface evaporation. On the bottom panel, the top abscissa shows the diagnosed Brunt--V\"ais\"al\"a frequency \(N\) obtained from the potential temperature \(\theta = T\,(p_0/p)^{R/c_p}\) where $p_0$ is a reference pressure and $R$ is the specific gas constant. Background fills indicate the sign of \(N^2=\frac{g}{\theta}\,\frac{d\theta}{dz}\) (stable, neutral, or unstable); superposed diagonal hatching partitions the stable column into three stratification regimes used below: \emph{soft} (\(N<0.010~\mathrm{s^{-1}}\)), \emph{moderate} (\(0.010\le N<0.015~\mathrm{s^{-1}}\)), and \emph{stiff} (\(N\ge 0.015~\mathrm{s^{-1}}\)). The present CO$_2$ column is everywhere statically stable (\(N^2>0\)), but with substantial vertical variation in \(N\): values are relatively large near the surface, dip to a broad mid-column minimum (\(\sim\!10^{-2}~\mathrm{s^{-1}}\)), and strengthen again aloft.

\subsubsection{NBH and Overshoot at TRAPPIST-1e moist terminator}\label{subsub:stab}

\begin{figure}
  \centering
  \includegraphics[width=0.78\textwidth]{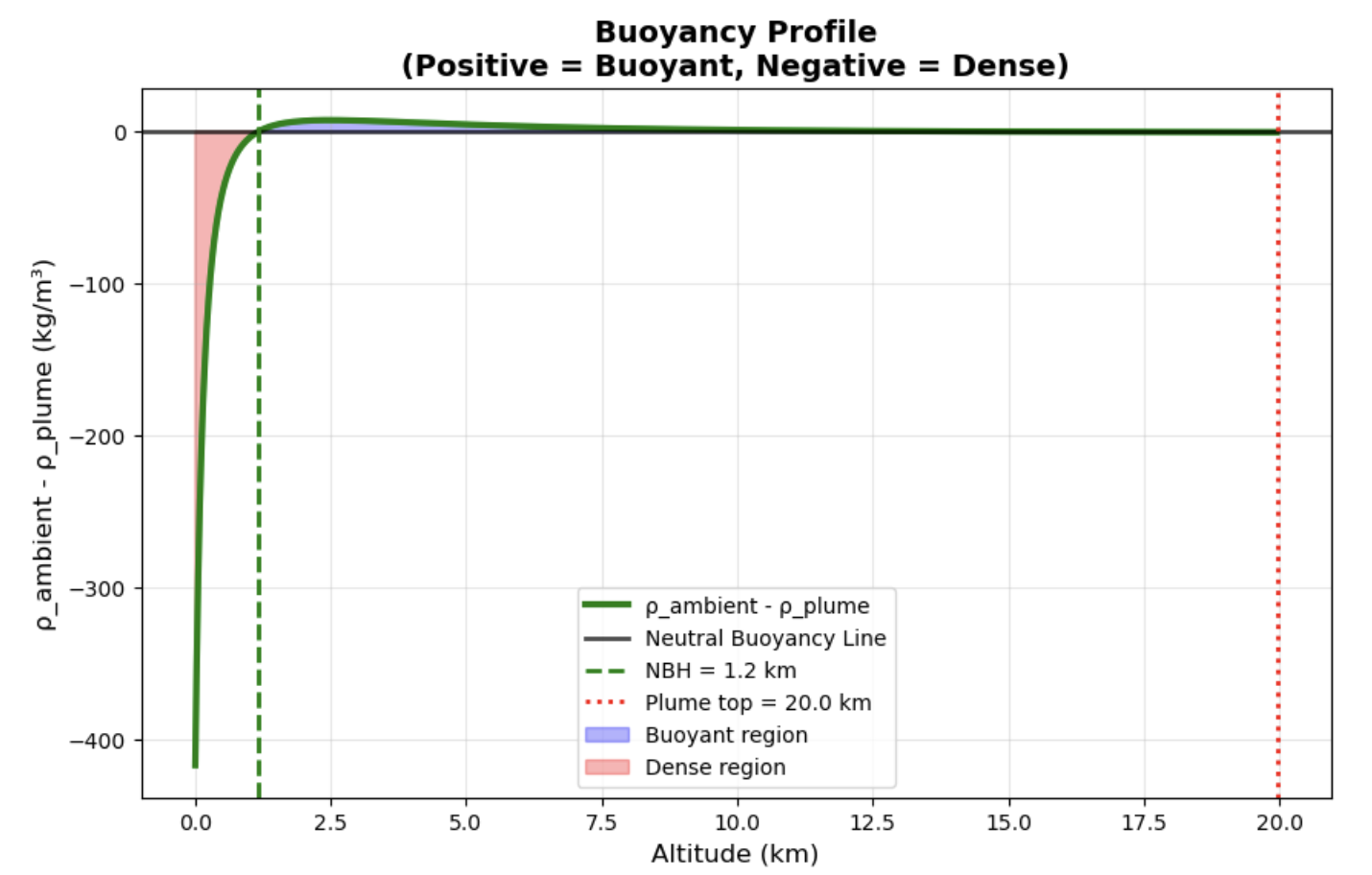}
  \caption{Plume buoyancy diagnostic for our TRAPPIST-1e, 2~bar CO$_2$ single column. Centerline density contrast
  \(\rho_a-\rho_p\) (green), neutral-buoyancy line (black), diagnosed NBH
  at \(z_{\rm NBH}\simeq 1.2~\mathrm{km}\) (green dashed), and plume-top
  \(H\simeq 20~\mathrm{km}\) (red dotted). Above NBH the contrast is weakly
  positive, and the large overshoot \(\Delta z\) is explained by finite
  momentum at NBH rising through a \emph{soft} mid-column, followed by arrest
  in a \emph{stiff} upper layer (see Fig.~\ref{fig:tpn}).}
  \label{fig:buoy_diag}
\end{figure}

Figure~\ref{fig:buoy_diag} shows the corresponding plume buoyancy diagnostic: the centerline density contrast \(\rho_a(z)-\rho_p(z)\) where $\rho_a$ is ambient density and $\rho_p$ is plume density, the neutral-buoyancy height \(z_{\rm NBH}\simeq 1.2~\mathrm{km}\) where $\rho_a = \rho_p$, and the diagnosed plume-top \(H\simeq 20~\mathrm{km}\). The plume is initialized as a high-momentum jet (\(\sim\!250~\mathrm{m\,s^{-1}}\) at the vent), entrains rapidly, crosses to positive buoyancy at NBH, and then rises through layers where \(\rho_a-\rho_p\) is only weakly positive. The large separation \(\Delta z \equiv H-z_{\rm NBH}\simeq 18.8~\mathrm{km}\) is therefore an \emph{overshoot} driven by finite vertical momentum into a stratification that is weak over much of the ascent and stiffens aloft.

A constant-\(N\) estimate \(H \approx z_{\rm NBH}+w_{\rm NBH}/N\) where $w_{\rm NBH}$ is the vertical velocity at neutral buoyancy height is not appropriate for this column because \(N\) varies significantly with height. Instead, we use an energy balance with vertically varying stability,
\begin{equation}
\frac{1}{2}\,w_{\rm NBH}^{2}
\;\approx\;
\int_{z_{\rm NBH}}^{H} N^{2}(z)\,\bigl(z-z_{\rm NBH}\bigr)\,dz
\;\equiv\;\frac{1}{2}\,N_{\rm eff}^{2}\,\Delta z^{2},
\label{eq:energy_overshoot}
\end{equation}
which defines an energy-weighted effective stability \(N_{\rm eff}\) for the layer actually traversed by the overshoot. Using the diagnosed \(N(z)\) from Fig.~\ref{fig:tpn} and the measured \(\Delta z\), we obtain \(N_{\rm eff}\approx 1.17\times 10^{-2}~\mathrm{s^{-1}}\) and hence
\begin{equation}
w_{\rm NBH}\;\approx\;N_{\rm eff}\,\Delta z
\;\simeq\;2.2\times 10^{2}\ \mathrm{m\,s^{-1}}.
\label{eq:wnbh}
\end{equation}
This \(w_{\rm NBH}\) is physically consistent with a \(\sim\!250~\mathrm{m\,s^{-1}}\) source that decelerates by entrainment/drag before reaching NBH. The two-panel diagnostics are mutually consistent: the plume rises nearly ballistically through the \emph{soft} mid-column (small \(N\)) and is ultimately arrested as it encounters the \emph{stiff} upper layer (larger \(N\)), where the work against buoyancy becomes prohibitive.

Two practical sensitivities follow directly from eq.~\eqref{eq:energy_overshoot}. First, \(\partial H/\partial w_{\rm NBH}\approx N_{\rm eff}^{-1}\): for the present column, a \(+30~\mathrm{m\,s^{-1}}\) change in \(w_{\rm NBH}\) shifts \(H\) by \(\sim\!2.6~\mathrm{km}\). Second, perturbations that deepen or thicken the \emph{soft} region (e.g., warmer lapse rates) reduce \(N_{\rm eff}\) and increase \(\Delta z\) at fixed \(w_{\rm NBH}\), whereas stronger upper-level stability increases \(N_{\rm eff}\) and lowers \(H\).

\subsection{Atmospheric Instability Effects}\label{subsec:instability}

To probe plume dynamics across an \emph{unstable} layer ($N^2<0$) we deliberately maximized near-surface superadiabaticity in the ambient column. We simulated TRAPPIST-1d (closer/warmer dayside than TRAPPIST-1e) with ExoCAM for a dry \emph{land}-planet setup (removing water vapor that would absorb shortwave aloft and strongly stabilize the lapse rate) with zero topography, and a thinner 0.5~bar CO$_2$ atmosphere (shorter radiative timescales and steeper daytime surface heating). We extracted the \emph{substellar} column (peak insolation) where the surface energy input is largest.

\subsubsection{TRAPPIST-1d Atmospheric Profile}

\begin{figure}
  \centering
  \includegraphics[width=0.5\textwidth]{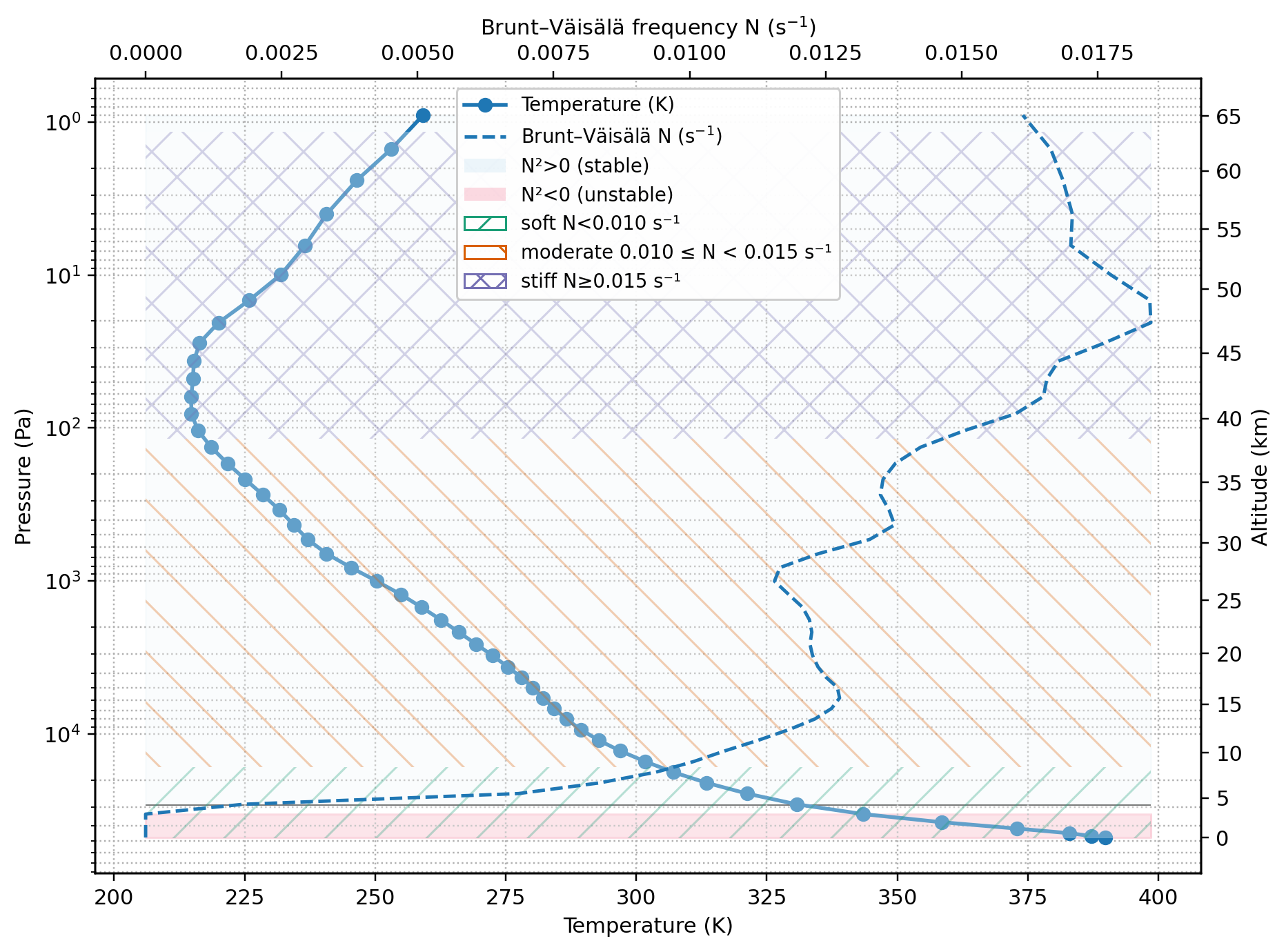}
  \caption{Similar figure to the bottom panel of Fig.~\ref{fig:tpn} but for TRAPPIST-1d substellar point with a dry 0.5~bar CO$_2$ atmosphere. The atmosphere clearly shows an unstable region (red shade) in the first $\sim$4~km.}
  \label{fig:tpnt1d}
\end{figure}

The profile is shown in Fig.~\ref{fig:tpnt1d}. Similarly to Fig.~\ref{fig:tpn}, the top abscissa displays the diagnosed Brunt--V\"ais\"al\"a frequency \(N\) derived from the value of the potential temperature \(\theta = T\,(p_0/p)^{R/c_p}\). Background fills indicate the sign of $N^2$ (stable, neutral, or unstable); superposed diagonal hatching partitions the stable column into the three stratification regimes previously defined.

In this new configuration, the diagnosed potential-temperature gradient $d\theta/dz$ becomes negative over roughly the first $\sim$4~km, so
\[
N^2(z) \;=\; \frac{g}{\theta}\,\frac{d\theta}{dz} \;<\; 0,
\qquad
\theta \;=\; T\!\left(\frac{p_0}{p}\right)^{R/c_p}, 
\]
with $g\simeq 6.11~\mathrm{m\,s^{-2}}$ (TRAPPIST-1d), $c_p=846~\mathrm{J\,kg^{-1}\,K^{-1}}$, $R=188.9~\mathrm{J\,kg^{-1}\,K^{-1}}$, implying a dry-adiabatic lapse $\Gamma_d=g/c_p\simeq 7.2~\mathrm{K\,km^{-1}}$. When the environmental lapse exceeds $\Gamma_d$, $\theta$ decreases with $z$ and parcels experience buoyant growth with e-folding time $\tau_c\simeq 1/|N|$. Consequently, a ``hot-balloon'' test (zero initial velocity) should \emph{still} accelerate upward through the $N^2<0$ layer; a momentum-driven plume will emerge from the top of the unstable layer with additional vertical kinetic energy, after which its rise is governed by the overlying $N^2>0$ stratification (e.g., an energy-weighted $N_{\rm eff}$ for overshoot).

\subsubsection{Atmospheric Instability Effects: Stable vs Unstable Stratification}
\label{subsec:instability}

To isolate the impact of atmospheric stability on volcanic plume dynamics, we conducted low-velocity ``hot air balloon'' tests designed to minimize initial momentum effects and emphasize buoyancy-driven ascent. These experiments use initial velocities of $10~{\rm m\,s^{-1}}$, which are substantially lower than the $250~{\rm m\,s^{-1}}$ typical of momentum-dominated VEI4 eruptions, allowing us to examine how plumes rise through atmospheres with different stability profiles when buoyancy forces dominate over initial kinetic energy.  We note that in momentum dominated plumes with higher initial velocities, the impact of this instability is muted as it becomes a second order effect.

The high water mass fractions ($n = 0.18$--0.30) employed in these tests are necessary to achieve near-buoyant conditions in the dense 0.5 bar CO$_2$ atmosphere, where replacing heavy solid particles with lighter water vapor is required to reduce plume density sufficiently for buoyancy-driven rise. This contrasts with Earth volcanism, where lower water contents (more typical of observed eruptions) can produce substantial plume rise through initial momentum supplied by high exit velocities, even when plumes are initially denser than the surrounding atmosphere. The low-velocity, high-water-content experimental design thus isolates the buoyancy physics relevant to volcanic injection in dense exoplanet atmospheres.

The controlled comparison between stable (TRAPPIST-1e) and unstable (TRAPPIST-1d) atmospheric profiles at identical 0.5 bar CO$_2$ surface pressure reveals that unstable stratification significantly enhances volcanic injection heights once buoyancy thresholds are achieved, representing a 2.5$\times$ height increase over stable conditions (Table~\ref{tab:stable_unstable_comparison}). This difference in how volcanic plumes interact with atmospheric stability has implications for understanding volcanic climate impacts on M-dwarf exoplanets. The effect demonstrates that atmospheric instability acts as an amplifier of buoyancy-driven ascent rather than simply lowering buoyancy requirements, creating sharp injection thresholds where plumes either fail completely or achieve substantial injection heights with limited intermediate behavior.

\begin{table}[h!]
\centering
\caption{Stable vs Unstable Atmosphere Plume Heights at 0.5 bar CO$_2$ (Low-Velocity Tests, $w_0 = 10~{\rm m\,s^{-1}}$)}
\begin{tabular}{|c|c|c|}
\hline
\textbf{Water Mass} & \textbf{TRAPPIST-1e} & \textbf{TRAPPIST-1d} \\
\textbf{Fraction ($n$)} & \textbf{Stable (km)} & \textbf{Unstable (km)} \\
\hline
0.03 & 0.005 & 0.005 \\
0.10 & 0.005 & 0.005 \\
0.18 & 4.7 & 0.005 \\
0.19 & 4.7 & 11.6 \\
0.20 & 4.6 & 11.6 \\
0.30 & 4.2 & 11.4 \\
\hline
\end{tabular}
\label{tab:stable_unstable_comparison}
\end{table}

For TRAPPIST-1d unstable atmosphere (Table~\ref{tab:stable_unstable_comparison}), complete injection failure occurs at water mass fractions $n=0.03, 0.10, 0.18$ (where $n$ is the ratio of water mass to total magmatic mass), with plumes reaching only 5 meters height. However, at $n=0.19$, plumes achieve injection heights of 11.6 km, maintained at 11.6 km for $n=0.20$ and declining slightly to 11.4 km at $n=0.30$. This represents a sharp transition where plumes either fail to rise or achieve near-maximum injection efficiency with minimal intermediate states.

In contrast, TRAPPIST-1e stable atmosphere demonstrates more gradual threshold behavior. Complete failure occurs at $n=0.03$ and $n=0.10$, but partial success begins at $n=0.18$ with 4.7 km injection height. Height remains relatively stable across higher water fractions: 4.7 km ($n=0.19$), 4.6 km ($n=0.20$), declining to 4.2 km ($n=0.30$). The stable atmosphere exhibits a more gradual response with moderate but consistent injection heights above threshold.

As shown in Table~\ref{tab:stable_unstable_comparison}, the unstable atmosphere provides a 2.5$\times$ height enhancement over stable conditions (11.6 km vs 4.7 km at $n=0.20$), representing substantial amplification of volcanic injection efficiency. The low-velocity experimental design ensures this enhancement reflects atmospheric instability effects on buoyancy-driven ascent rather than momentum-related dynamics. This demonstrates that atmospheric instability amplifies buoyancy-driven ascent rather than simply lowering threshold requirements.

\subsubsection{Isolation of Gravity and Instability Effects}

To separate atmospheric instability effects from gravitational differences between TRAPPIST-1d ($g = 6.12$ m/s$^2$) and TRAPPIST-1e ($g = 8.01$ m/s$^2$), we conducted controlled experiments using TRAPPIST-1d atmospheric conditions with TRAPPIST-1e gravitational acceleration. At $n=0.18$, complete failure persists regardless of gravity, confirming the atmospheric threshold effect. At $n=0.20$, TRAPPIST-1d atmospheric conditions with TRAPPIST-1e gravity achieve 8.9 km compared to 11.6 km under TRAPPIST-1d gravity, indicating that atmospheric instability provides the primary amplification ($\sim$2$\times$ over stable conditions) with reduced gravity contributing additional enhancement ($\sim$1.3$\times$).

The combined effects are not simply multiplicative: atmospheric instability enables the fundamental transition from failure to success, while lower gravity enhances the magnitude of successful injection. This suggests that M-dwarf exoplanets with unstable surface layers could experience dramatically enhanced volcanic-atmospheric coupling compared to stable atmospheric analogs for buoyancy driven low velocity plumes, with implications for volatile cycling and atmospheric evolution on tidally heated worlds.

Note that the standard dry stability diagnostic, \(N^2 = \dfrac{g}{\theta}\,\dfrac{d\theta}{dz}\), assumes a hydrostatic, single–constituent atmosphere and small, adiabatic perturbations. Two important caveats apply: composition gradients can introduce Ledoux terms that modify the buoyancy frequency, and non-hydrostatic or strongly diabatic regimes require generalized stability diagnostics beyond the standard dry formulation.

\subsection{Surface Gravity Sensitivity Test}\label{subsec:gravity}

Surface gravity is a primary control on both the buoyancy that drives volcanic plumes and the stratification that resists their vertical motion. In the simplified scaling used here, the neutral-buoyancy and overshoot heights are linked through the vertically varying Brunt--V\"ais\"al\"a frequency via the energy balance in Equation \ref{eq:energy_overshoot}. To leading order, both the driving buoyancy and the stabilizing $N$ scale with $g$. For a fixed source temperature contrast, composition, and lapse rate, increasing $g$ strengthens the buoyancy force $g\,\Delta\rho/\rho$ but also steepens the restoring force through $N^2 \propto g\,\partial\theta/\partial z$, so that $N_{\rm eff}$ typically grows faster than $w_{\rm NBH}$. In this regime, Eq.~\eqref{eq:energy_overshoot} implies a net reduction of $\Delta z$ with increasing $g$, which leads to the intuitive result that plumes on higher-gravity planets will not be as high, when all other variables are controlled.

We conducted systematic gravity sensitivity experiments spanning surface accelerations from $1.5~{\rm m\,s^{-2}}$ (comparable to low-mass rocky planets) to $35~{\rm m\,s^{-2}}$ (extreme super-Earth conditions) using the baseline 2-bar CO$_2$ atmospheric profile. The results demonstrate systematic decline in plume injection heights with increasing gravity across both eruption intensities.

For VEI4 eruptions shown in table \ref{tab:gravity_effects}, plume heights decrease from the top of our simulated atmospheres at 61.5 km at the lowest gravity values ($g = 1.5$--$1.62~{\rm m\,s^{-2}}$) to only 2.2 km under extreme gravity ($g = 35~{\rm m\,s^{-2}}$). The decline is particularly pronounced between Mars-like gravity ($g = 3.7~{\rm m\,s^{-2}}$, $H = 54.4$ km) and Earth-like conditions ($g = 9.8~{\rm m\,s^{-2}}$, $H = 16.4$ km), representing a factor of 3.3 reduction in injection height. TRAPPIST-1 planet conditions ($g = 6.5$--$8.9~{\rm m\,s^{-2}}$) yield intermediate performance with heights of 18.8--26.7 km.

VEI6 eruptions exhibit enhanced resilience to gravitational effects due to their higher initial momentum, maintaining maximum injection heights of 61.5 km across the lowest gravity range ($g = 1.5$--$3.7~{\rm m\,s^{-2}}$) before declining to 40.1 km at Earth gravity and 3.8 km under extreme conditions. The transition occurs more gradually for VEI6, with substantial injection heights maintained until $g \gtrsim 15~{\rm m\,s^{-2}}$.

\begin{table}[h!]
\centering
\caption{Surface Gravity Effects on Volcanic Plume Heights}
\begin{tabular}{|l|c|c|c|}
\hline
\textbf{Gravity} & \textbf{Representative} & \textbf{VEI4 Height} & \textbf{VEI6 Height} \\
\textbf{(${\rm m\,s^{-2}}$)} & \textbf{Body} & \textbf{(km)} & \textbf{(km)} \\
\hline
1.5 & Low-mass rocky & 61.5 & 61.5 \\
1.62 & Moon & 61.5 & 61.5 \\
3.7 & Mercury/Mars & 54.4 & 61.5 \\
6.5 & TRAPPIST-1 range & 26.7 & 60.0 \\
8.9 & Venus & 18.8 & 44.4 \\
9.8 & Earth & 16.4 & 40.1 \\
12 & Super-Earth & 12.5 & 32.2 \\
20 & High-gravity & 7.4 & 16.1 \\
35 & Extreme gravity & 2.2 & 3.8 \\
\hline
\end{tabular}
\label{tab:gravity_effects}
\end{table}

Our sensitivity experiments confirm the qualitative expectation. When we vary $g$ at fixed vent conditions, injection speed, and background CO$_2$ column, the plume tops tend to reach similar pressure levels but different geometric altitudes: lower gravity yields taller plumes in kilometers, while higher gravity compresses the atmospheric scale height and reduces the overshoot. The dependence is strong and systematic, directly relevant for detectability. On sub-Earth-gravity worlds, plumes are more likely to penetrate to low pressures where they can imprint strong signatures in transmission or emission. Conversely, on high-gravity super-Earths, comparable source conditions produce shorter plumes that remain deeper in the atmosphere, reducing their leverage on the observable limb but potentially enhancing local radiative and chemical impacts near the surface.

\subsection{Pressure Effects and Optimal Atmospheric Density}
\label{subsec:pressure_effects}

The pressure series reveals non-monotonic injection efficiency with peak heights around 2-4 bar, declining at both lower and higher pressures. This behavior identifies an optimal atmospheric pressure window where dense atmospheres provide sufficient buoyancy for volcanic injection while entrainment penalties remain manageable. Below this window, atmospheres are insufficient to enable buoyancy, while above 64 bar, entrainment of extremely dense atmospheres creates prohibitive momentum decay that limits injection heights despite favorable buoyancy conditions. This suggests that volcanic-atmospheric coupling efficiency depends critically on finding intermediate pressure regimes that balance buoyancy enablement with entrainment costs.

\subsubsection{Uniform Pressure Scaling at Fixed Temperature}

We assess how the maximum plume rise height, $H$, responds to a uniform scaling of the \emph{ambient pressure profile} while holding the temperature profile from Fig.~\ref{fig:tpn} fixed. This experiment probes the roles of ambient density and static stability independent of thermal stratification changes. 

In our implementation, the vent boundary condition prescribes \emph{velocity} (not pressure), so the near-vent mass flux and thermodynamic state respond consistently to the local solution, and nozzle choking constraints at the boundary are not invoked.

Writing potential temperature $\theta(z)=T(z)\,[p_0/p(z)]^{\kappa}$ with $\kappa\equiv R/c_p$, a multiplicative factor $f$ in $p(z)$ multiplies both $\theta$ and $d\theta/dz$ by $f^{-\kappa}$, so
\begin{equation}
N^2 \;=\; \frac{g}{\theta}\,\frac{d\theta}{dz}
\;\;\;\Rightarrow\;\;\; N \ \text{is unchanged by}\ p\mapsto f\,p \ \text{when}\ T(z)\ \text{is fixed}.
\label{eq:N_invariance}
\end{equation}

\begin{figure}
\centering 
\includegraphics[width=0.7\columnwidth]{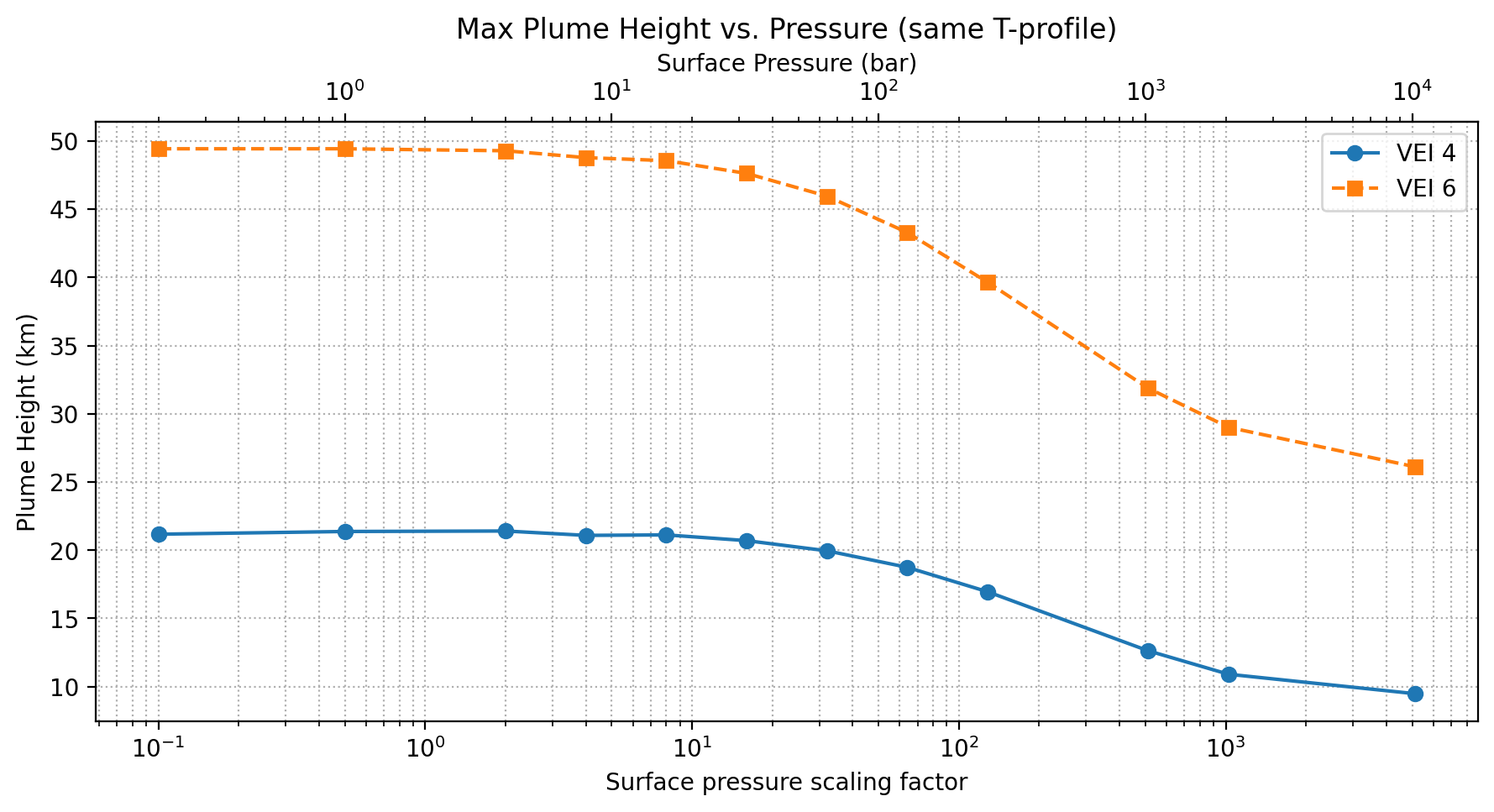}
\caption{Plume height $H$ dependency on surface pressure for VEI--4 and VEI--6. The temperature profile of Fig.~\ref{fig:tpn} is kept fixed but the pressure, from the surface to the top of the model is scaled by the $f$ factor of the bottom x-axis. The top x-axis shows the actual surface pressure values in bar.}
\label{Fig:pressure}
\end{figure}

Equations~(\ref{eq:F0}) and (\ref{eq:N_invariance}) therefore suggest that, to first order, $H$ should be nearly invariant under pressure scaling when $w_0$ (initial velocity) and $T(z)$ are fixed. Figure~\ref{Fig:pressure} shows the modeled response using the pressure scaling factor $f$ relative to a 2~bar surface pressure (i.e., $p_{\rm s}=2f$~bar). Both VEI--4 and VEI--6 exhibit near-constant $H$ for $f\lesssim 16$ (2--32~bar) with a gradual decline, followed by a stronger reduction for $f\ge 32$ (64--$10^4$~bar), consistent with increasing non-Boussinesq/compressible damping of the rise and enhanced entrainment cooling at extreme density.

\subsubsection{Non-monotonic Injection and Optimal Density Window}

A more realistic study of pressure effects with corresponding temperature variations necessitates consideration of underlying physics that involves dynamic feedback between three coupled processes: momentum decay through entrainment, thermodynamic evolution of multi-component plumes, and buoyancy transitions. As momentum-driven plumes entrain heavy CO$_2$ (molecular weight $44~{\rm g\,mol^{-1}}$), they experience rapid mass addition that accelerates momentum decay, but the thermodynamic consequences depend critically on atmospheric pressure and temperature profiles. Dense atmospheres enable buoyancy for solids-rich plumes, which are impossible in Earth-like conditions, but create vigorous entrainment that can overwhelm momentum before optimal heights are reached.

The optimal pressure window around 4 bar for low-momentum plumes (compared to 2 bar for high-momentum VEI4 eruptions) reflects a fundamental shift in the momentum-buoyancy transition point. Lower initial momentum requires denser ambient conditions to achieve buoyancy, but excessive density triggers entrainment feedback loops for the plume composition examined that truncate rise before momentum is fully converted to altitude. This creates a narrow pressure range where thermodynamic evolution pathways allow maximum momentum-to-altitude conversion efficiency.

\begin{table}[h!]
\centering
\caption{Pressure Effects on Low-Momentum Volcanic Plume Heights ($n=0.2$)}
\begin{tabular}{|c|c|c|c|c|c|}
\hline
\textbf{Pressure} & \textbf{(bar)} & \textbf{NBH (km)} & \textbf{Final Height (km)} & \textbf{Mass Multiplication} & \textbf{Overshoot (km)} \\
\hline
2 & Baseline & 4.4 & 5.7 & 83.5$\times$ & 1.3 \\
4 & Optimal & 5.4 & 7.2 & 93.5$\times$ & 1.8 \\
8 & Declining & 4.6 & 5.7 & 65.1$\times$ & 1.2 \\
\hline
\end{tabular}
\label{tab:pressure_effects_hotairballoon}
\end{table}

\subsection{Atmospheric Molecular Weight Effects}\label{subsec:mmw_effects}

The entrainment-driven dynamics observed in the pressure sensitivity tests depend fundamentally on the molecular weight contrast between volcanic volatiles and ambient atmospheric gases. To isolate these compositional effects, we conducted systematic tests across atmospheric molecular weights ranging from hydrogen (MMW=$2~{\rm g\,mol^{-1}}$) to krypton (MMW=$84~{\rm g\,mol^{-1}}$) under identical 2-bar pressure conditions (with n=0.03 for water abundance by mass), revealing complex non-monotonic behavior in injection efficiency.

The results reveal that pure hydrogen atmospheres (MMW=$2~{\rm g\,mol^{-1}}$) produce low injection heights of only 4.3 km for VEI4 eruptions, likely due to insufficient density contrast for effective buoyancy despite the light ambient gas. However, high-metallicity hydrogen atmospheres (MMW=$8~{\rm g\,mol^{-1}}$) achieve the maximum injection heights observed in our study: 32.3 km for VEI4 and 74.4 km for VEI6 eruptions, representing optimal conditions where moderate atmospheric density provides buoyancy while minimizing entrainment penalties.

Intermediate molecular weight atmospheres show systematic decline with increasing density: methane (MMW=$16~{\rm g\,mol^{-1}}$) achieves 26.7 km, water vapor (MMW=$18~{\rm g\,mol^{-1}}$) reaches 25.9 km, and Earth-like nitrogen atmospheres (MMW=$29~{\rm g\,mol^{-1}}$) produce 23.4 km injection heights for VEI4 eruptions. Standard CO$_2$ atmospheres (MMW=$44~{\rm g\,mol^{-1}}$) yield 21.5 km, while heavier compositions show continued decline: SO$_2$ atmospheres (MMW=$64~{\rm g\,mol^{-1}}$) reach 19.6 km, iron oxide (MMW=$72~{\rm g\,mol^{-1}}$) produces 19.0 km, and krypton (MMW=$84~{\rm g\,mol^{-1}}$) achieves 18.1 km for VEI4 cases.

The molecular weight dependence becomes more pronounced for higher-energy VEI6 eruptions, which achieve 49.7 km in CO$_2$ versus 59.3 km in water vapor atmospheres, demonstrating that higher initial momentum can better overcome entrainment penalties in dense atmospheric compositions. Notably, VEI6 eruptions in the lightest atmospheres (methane, water vapor) achieve injection heights approaching 60 km, nearly matching the theoretical maximum (though at these lower pressures, we note that the trend is more important than specific numbers) observed at very low gravity conditions.

\begin{table}[h!]
\centering
\caption{Atmospheric Molecular Weight Effects on Volcanic Plume Heights}
\begin{tabular}{|l|c|c|c|}
\hline
\textbf{Atmospheric} & \textbf{MMW} & \textbf{VEI4 Height} & \textbf{VEI6 Height} \\
\textbf{Composition} & \textbf{(${\rm g\,mol^{-1}}$)} & \textbf{(km)} & \textbf{(km)} \\
\hline
Hydrogen & 2 & 4.3 & 10.6 \\
High-metallicity H$_2$ & 8 & 32.3 & 74.4 \\
Methane & 16 & 26.7 & 60.1 \\
Water vapor & 18 & 25.9 & 59.3 \\
Earth/Nitrogen & 29 & 23.4 & 54.3 \\
Carbon Dioxide & 44 & 21.5 & 49.7 \\
Sulfur Dioxide & 64 & 19.6 & 45.0 \\
Iron Oxide & 72 & 19.0 & 43.7 \\
Krypton & 84 & 18.1 & 42.0 \\
\hline
\end{tabular}
\label{tab:mmw_effects}
\end{table}

These molecular weight effects reveal that exoplanet atmospheric composition represents a first-order control on volcanic injection efficiency. The results demonstrate a fundamental atmospheric threshold around MMW=$8$--$16~{\rm g\,mol^{-1}}$ for the cases we tested, where optimal injection occurs, followed by systematic decline as atmospheric molecular weight increases. The substantial molecular weight difference between volcanic volatiles (primarily H$_2$O with MMW=$18~{\rm g\,mol^{-1}}$) and heavy atmospheric gases (CO$_2$ with MMW=$44~{\rm g\,mol^{-1}}$) creates strong entrainment penalties that fundamentally alter plume dynamics compared to Earth-like conditions.

\subsection{Water Content vs Plume Height}\label{subsec:water_content}

The molecular weight contrast identified in the previous section has profound implications for how water content affects plume injection heights. As a consequence, we conducted sensitivity tests that examined the impact of varying water mass fraction in the plume on plume height. In TRAPPIST-1e's dense CO$_2$ atmosphere, an inverse relationship emerges between water content and plume height, driven by a cascade of physical processes initiated by volume-constrained initial conditions. Since vent radius (100 m) and velocity ($10~{\rm m\,s^{-1}}$) are held constant, the volumetric flux $Q_0 = \pi r_0^2 w_0$ remains fixed while different water mass fractions create different vent mixture densities. Higher water content replaces dense solid particles ($1000~{\rm kg\,m^{-3}}$) with lighter water vapor, reducing the total initial mass flux and creating stronger initial buoyancy in the dense CO$_2$ atmosphere.

This enhanced buoyancy drives more vigorous entrainment of dense CO$_2$ (molecular weight $44~{\rm g\,mol^{-1}}$) compared to the lighter water vapor ($18~{\rm g\,mol^{-1}}$) being released. The substantial molecular weight difference means each unit of entrained CO$_2$ adds significantly more mass than the water vapor it mixes with, leading to rapid plume densification. Higher water content plumes therefore reach their neutral buoyancy height earlier and at lower altitudes, having less ability to build momentum before transitioning to the overshoot phase.

This mechanism contrasts sharply with Earth's atmosphere, where the entrainment penalty is much smaller due to air's lower molecular weight ($\sim$$29~{\rm g\,mol^{-1}}$ versus $44~{\rm g\,mol^{-1}}$ for CO$_2$), allowing positive effects of water content such as latent heat release and convective enhancement to play a more significant role. The findings suggest that atmospheric molecular weight may represent a critical threshold determining whether water content enhances or diminishes volcanic plume heights on different worlds.

\begin{table}[h!]
\centering
\caption{Volcanic Plume Characteristics vs Water Mass Fraction in TRAPPIST-1e 2-bar CO$_2$ Atmosphere}
\begin{tabular}{|c|c|c|c|c|c|c|}
\hline
\textbf{Water} & \textbf{NBH} & \textbf{Final Height} & \textbf{Initial Mass Flux} & \textbf{Final Mass Flux} & \textbf{Mass} & \textbf{CO$_2$ Entrainment} \\
\textbf{Fraction} & \textbf{(km)} & \textbf{(km)} & \textbf{($\times 10^5~{\rm kg\,s^{-1}}$)} & \textbf{($\times 10^7~{\rm kg\,s^{-1}}$)} & \textbf{Multiplication} & \textbf{Ratio} \\
\hline
0.18 & 4.5 & 5.8 & 2.2 & 2.1 & 78.2$\times$ & 75.6 \\
0.2 & 4.4 & 5.7 & 2.0 & 2.0 & 83.5$\times$ & 80.7 \\
0.3 & 4.1 & 5.3 & 1.1 & 1.8 & 109.3$\times$ & 106.0 \\
0.4 & 4.0 & 5.1 & 0.7 & 1.6 & 134.9$\times$ & 131.1 \\
0.5 & 3.9 & 5.0 & 0.5 & 1.6 & 160.3$\times$ & 156.0 \\
\hline
\end{tabular}
\label{tab:water_content_analysis}
\end{table}

Table~\ref{tab:water_content_analysis} clearly demonstrates the inverse relationship in a low momentum test that is buoyancy driven: as water mass fraction increases from 0.18 to 0.5, neutral buoyancy height decreases from 4.474 km to 3.891 km, and final plume height drops from 5.81 km to 5.0 km. Simultaneously, the initial mass flux decreases dramatically due to the replacement of dense solids with light water vapor, while the mass multiplication factor increases from 78$\times$ to 160$\times$, indicating progressively more vigorous CO$_2$ entrainment that drives the earlier neutral buoyancy and reduced final heights.

To verify that this counterintuitive relationship is driven by entrainment processes rather than temperature-dependent thermodynamic effects, we conducted isothermal atmospheric tests (Table~\ref{tab:isothermal_comparison}). These experiments eliminate all temperature stratification while maintaining identical pressure profiles, isolating the entrainment mechanism.

\begin{table}[h!]
\centering
\caption{Isothermal vs Stratified Atmosphere Comparison for Volcanic Plume Dynamics}
\begin{tabular}{|l|c|c|c|c|c|c|}
\hline
\textbf{Atmosphere} & \textbf{Water} & \textbf{Thermal} & \textbf{NBH} & \textbf{Final Height} & \textbf{Mass} & \textbf{CO$_2$ Entrainment} \\
\textbf{Type} & \textbf{Fraction} & \textbf{Contrast (K)} & \textbf{(km)} & \textbf{(km)} & \textbf{Multiplication} & \textbf{Ratio} \\
\hline
Stratified (321$\rightarrow$309K) & 0.2 & 579$\rightarrow$591 & 4.4 & 5.7 & 83.5$\times$ & 80.7 \\
Stratified (321$\rightarrow$309K) & 0.4 & 579$\rightarrow$591 & 4.0 & 5.1 & 134.9$\times$ & 131.1 \\
\hline
Isothermal 315K & 0.2 & 585 & 3.4 & 4.4 & 60.9$\times$ & 58.6 \\
Isothermal 315K & 0.4 & 585 & 3.1 & 4.0 & 100.5$\times$ & 97.5 \\
\hline
Isothermal 340K & 0.2 & 560 & 3.4 & 4.4 & 54.8$\times$ & 52.7 \\
Isothermal 340K & 0.4 & 560 & 3.1 & 4.0 & 93.1$\times$ & 90.1 \\
\hline
\end{tabular}
\label{tab:isothermal_comparison}
\end{table}

The isothermal atmospheric tests definitively confirm that the inverse water-height relationship is driven by entrainment processes rather than thermodynamic density evolution from temperature changes. In both 315K and 340K isothermal atmospheres, higher water content plumes consistently reach neutral buoyancy height earlier and achieve lower final altitudes compared to lower water content plumes, with the NBH differences ranging from 250-267 meters across temperature conditions. The relationship persists despite removing all temperature-induced thermodynamic effects and result in similar drops in neutral buoyancy and final plume heights to the stratified case, proving that differential CO$_2$ entrainment rates are the fundamental driver of the mechanism.

\subsection{Temperature Sensitivity Test}
\label{sec:temp_offsets_CO2}

The isothermal tests presented in Section~\ref{subsec:water_content} demonstrate that the inverse water-height relationship persists even when temperature stratification is eliminated, confirming entrainment as the fundamental mechanism rather than stratification-dependent behavior. However, temperature stratification does independently control injection thresholds through its influence on atmospheric stability, with isothermal atmospheres showing reduced injection heights compared to realistic stratified profiles. The stratification effects operate independently of the entrainment-driven water fraction responses, demonstrating that both atmospheric structure and composition must be considered when evaluating volcanic injection potential on exoplanets.

\begin{figure}[H]
\centering 
\includegraphics[width=0.7\columnwidth]{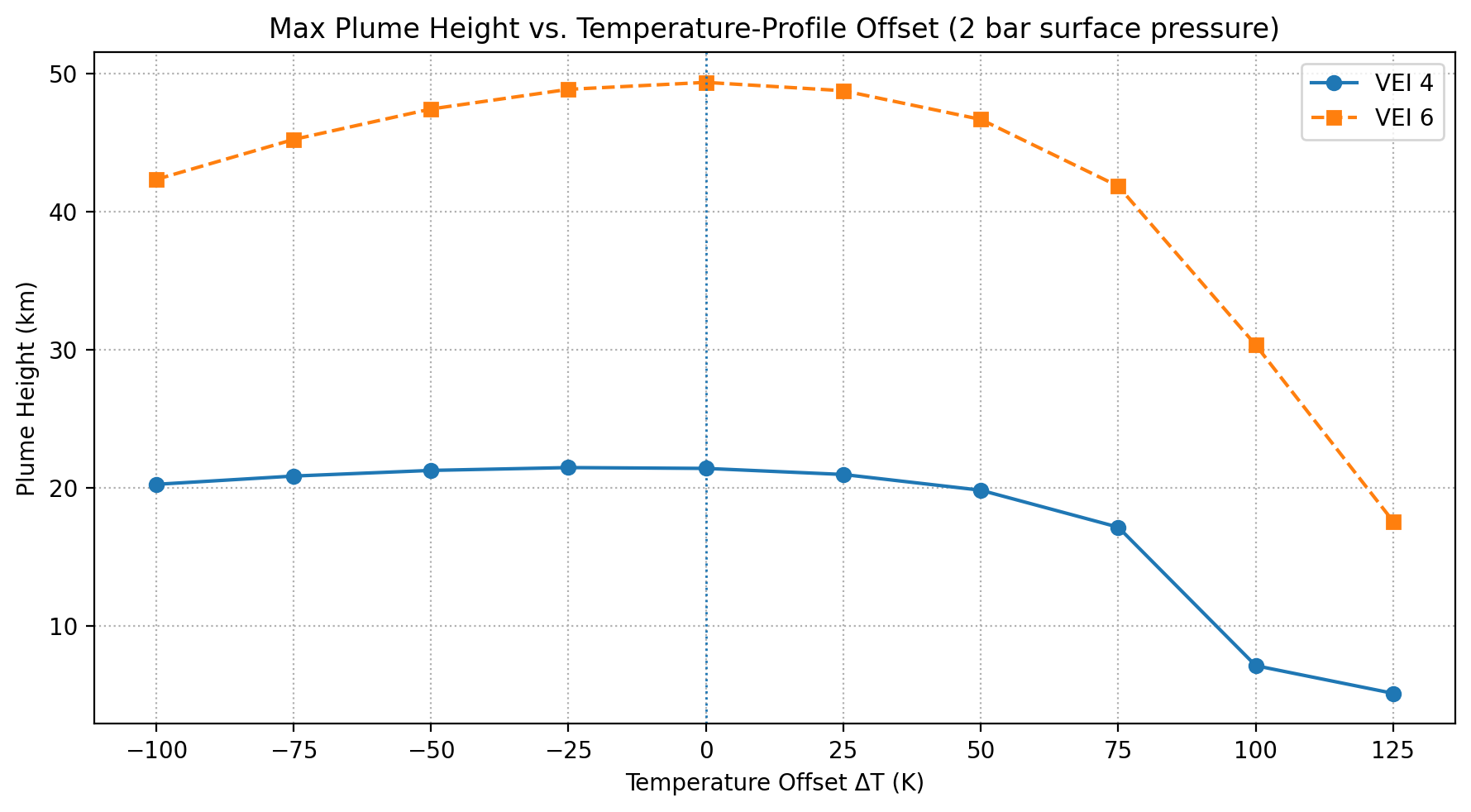}
\caption{Plume height $H$ dependency on atmospheric temperature for VEI--4 and VEI--6. The pressure profile of Fig.~\ref{fig:tpn} is kept unchanged but the temperature is changed, offset by a constant from the surface to the top of the model by a factor $\Delta T$ on the x-axis.}
\label{Fig:temperature}
\end{figure}

In order to assess the impact of the ambient air temperature alone on the plume buoyancy for a realistic, non-isothermal atmospheric column, we perturb the ambient temperature profile by a uniform offset $\Delta T_a$ while holding the 2~bar CO$_2$ pressure profile of Fig.~\ref{fig:tpn} (including the upper inversion) fixed. The vent boundary condition prescribes velocity, so the imposed source volume and momentum fluxes are held constant across offsets. Figure~\ref{Fig:temperature} reports the \emph{maximum} rise height $H_{\max}$ for VEI--4 ($T_p\!=\!900$~K) and VEI--6 ($T_p\!=\!1000$~K).

As previously detailed in Equations \ref{eq:H_scaling} to \ref{eq:N_iso}, first-order plume theory suggests $H \propto F_0^{1/4} N^{-3/4}$, where $F_0 \approx g Q_0(\Delta\rho/\rho_a)$ and, for fixed composition, $\Delta\rho/\rho_a \approx \Delta T/T_a$ at the vent, while $N^2 = \tfrac{g}{T}\!\left(\tfrac{dT}{dz} + \tfrac{g}{c_p}\right)$. A uniform offset $T \mapsto T+\Delta T_a$ leaves $dT/dz$ (the temperature gradient) unchanged but rescales $N^2 \propto 1/(T+\Delta T_a)$. Accordingly, warming reduces $F_0$ faster than it relieves stability ($N\!\downarrow$), and cooling increases $N$ and the density-weighted mass entrainment rate. Both effects drive $H_{\max}$ downward away from $\Delta T_a\!=\!0$, consistent with the simulations.

\subsection{Entrainment Parameter Sensitivity Test}\label{sec:entrain_test}

The entrainment parameter $\alpha$, which governs the rate at which ambient atmospheric material is incorporated into rising volcanic plumes, was originally calibrated by \citet{morton1956turbulent} using laboratory experiments in stratified salt solutions with Prandtl numbers around 80, yielding $\alpha \approx 0.093$. However, recent laboratory experiments demonstrate that entrainment coefficients vary significantly with atmospheric stratification, crossflow conditions, and plume dynamics \citep{aubry2017turbulent, carazzo2014laboratory}, while numerical studies reveal substantial sensitivity to atmospheric composition and density structure \citep{carazzo2008rise}. The appropriate entrainment coefficient for exoplanetary atmospheres likely differs significantly from this Earth-calibrated value due to variations in atmospheric composition, pressure, temperature, structure, and gravitational conditions that affect key fluid dynamic parameters including Prandtl, Richardson, and Reynolds numbers.

To quantify the impact of entrainment parameter uncertainty on exoplanetary volcanic plume heights, we conducted systematic sensitivity tests across $\alpha = 0.05$--0.17 for two contrasting atmospheric scenarios: moderate-pressure (1-bar) and high-pressure (8-bar) CO$_2$ atmospheres representing the range of conditions predicted for TRAPPIST-1e.  We examined a broader range than those typically incorporated in studies of solar system atmospheres to explore a wide range of potential scenarios that may be appropriate for the varied atmospheres expected on exoplanets.

The results demonstrate inverse dependence of plume height on entrainment parameter for moderate-pressure conditions. For 1-bar atmospheres, a decrease in $\alpha$ from 0.17 to 0.05 increases injection heights from 12.8 km to 21.8 km, representing a 70\% enhancement. This reflects the fundamental physics of entrainment whereby lower $\alpha$ values reduce the rate of ambient atmosphere incorporation, maintaining plume momentum longer and enabling greater altitude penetration before neutral buoyancy is reached.

The 8-bar high-pressure case reveals more complex behavior with an entrainment coefficient around $\alpha = 0.06$ that achieves maximum heights of 14.4 km. Both higher and lower $\alpha$ values produce reduced heights as $\alpha = 0.05$ yields only 9.2 km while $\alpha = 0.17$ produces 9.1 km. This non-monotonic response suggests that in dense atmospheres, there exists an optimal entrainment rate that balances momentum conservation against the benefits of thermal mixing and buoyancy enhancement from entrained ambient CO$_2$.

\begin{table}[h!]
\centering
\caption{Entrainment Parameter Effects on Volcanic Plume Heights in CO$_2$ Atmospheres}
\begin{tabular}{|c|c|c|}
\hline
\textbf{Entrainment} & \textbf{1-bar Height} & \textbf{8-bar Height} \\
\textbf{Parameter ($\alpha$)} & \textbf{(km)} & \textbf{(km)} \\
\hline
0.05 & 21.8 & 9.2 \\
0.08 & 17.5 & 13.6 \\
0.11 & 15.3 & 11.5 \\
0.14 & 13.9 & 10.1 \\
0.17 & 12.8 & 9.1 \\
\hline
\end{tabular}
\label{tab:entrainment_effects}
\end{table}

The pressure-dependent entrainment response shows that the entrainment coefficient that leads to maximum plume heights in a particular exoplanetary atmosphere varies significantly with atmospheric density. In moderate-pressure atmospheres, lower entrainment coefficients ($\alpha \approx 0.05$--0.07) appear to lead to the largest plume heights, while the maximum plumes heights in dense atmospheres correspond to around $\alpha \approx 0.06$, suggesting that entrainment efficiency varies with atmospheric density and composition in ways not captured by the standard Earth-calibrated value.

These findings highlight that volcanic plume height predictions for exoplanets carry significant uncertainty directly related to entrainment parameter choice. The factor-of-two height variations observed across the modeled $\alpha$ range represent uncertainty that may affect interpretations of volcanic injection efficiency and atmospheric impact assessments \citep{aubry_jellinek_condensation}. Future experimental validation or high-fidelity computational fluid dynamics studies \citep{suzuki2010numerical} specifically designed for non-terrestrial atmospheric conditions will be essential for constraining entrainment coefficients and reducing this uncertainty in exoplanet volcanic modeling.

\section{Discussion}
\label{sec:discussion}

Section \ref{sec:results} described the results of the various tests we ran in the context of the specific variables and processes we examined.  In this section we discuss implications of our work on other applicable topics, which we divide into subsections, though we broadly focus on observational implications and areas for future work, particularly in exploring underpinning geophysics and chemistry.

\subsection{Significant Potential Impacts in the Context of Model Assumptions and Understanding Limits}

While these results collectively suggest that M-dwarf exoplanets with unstable surface layers and intermediate atmospheric pressures ($\sim$2--4~bar) could experience mechanisms that enhance volcanic--atmospheric coupling for specific types of eruptions compared to stable atmospheric analogs, key context is required for a more comprehensive analysis of the impact of these results.  For example, while the identification of an optimal pressure window for volcanic injection indicates that the most significant volcanic climate impacts could occur on worlds within this intermediate pressure range, where atmospheric conditions maximize injection efficiency while maintaining sufficient density for substantial climate perturbation, this is dependent on accurate volcanic plume modeling and atmospheric modeling. Understanding limits given particular atmospheric and plume conditions \citep{glaze2002volcanic} will be key when applying such results to specific cases and may necessitate the use of more sophisticated models to capture potentially observable impacts.  Additionally, this window is not expected to be universal: it may vary between planetary systems and could shift over time as both atmospheric composition and volcanic activity patterns evolve during planetary evolution.

It should be noted that these simulations employ Earth-based volcanic plume parameterizations applied to exoplanet atmospheric conditions. This represents a reasonable starting point for exploring the relevant parameter space, while acknowledging that actual exoplanet volcanism may exhibit different characteristics (e.g., composition, conduit dynamics, fragmentation physics) that could modify the detailed scaling relationships.  

\subsection{An Example of Gravity Controls on Magma Volatiles, Eruption Style, and Injectability}

\begin{figure}[htbp]
    \centering
    \includegraphics[width=0.85\textwidth]{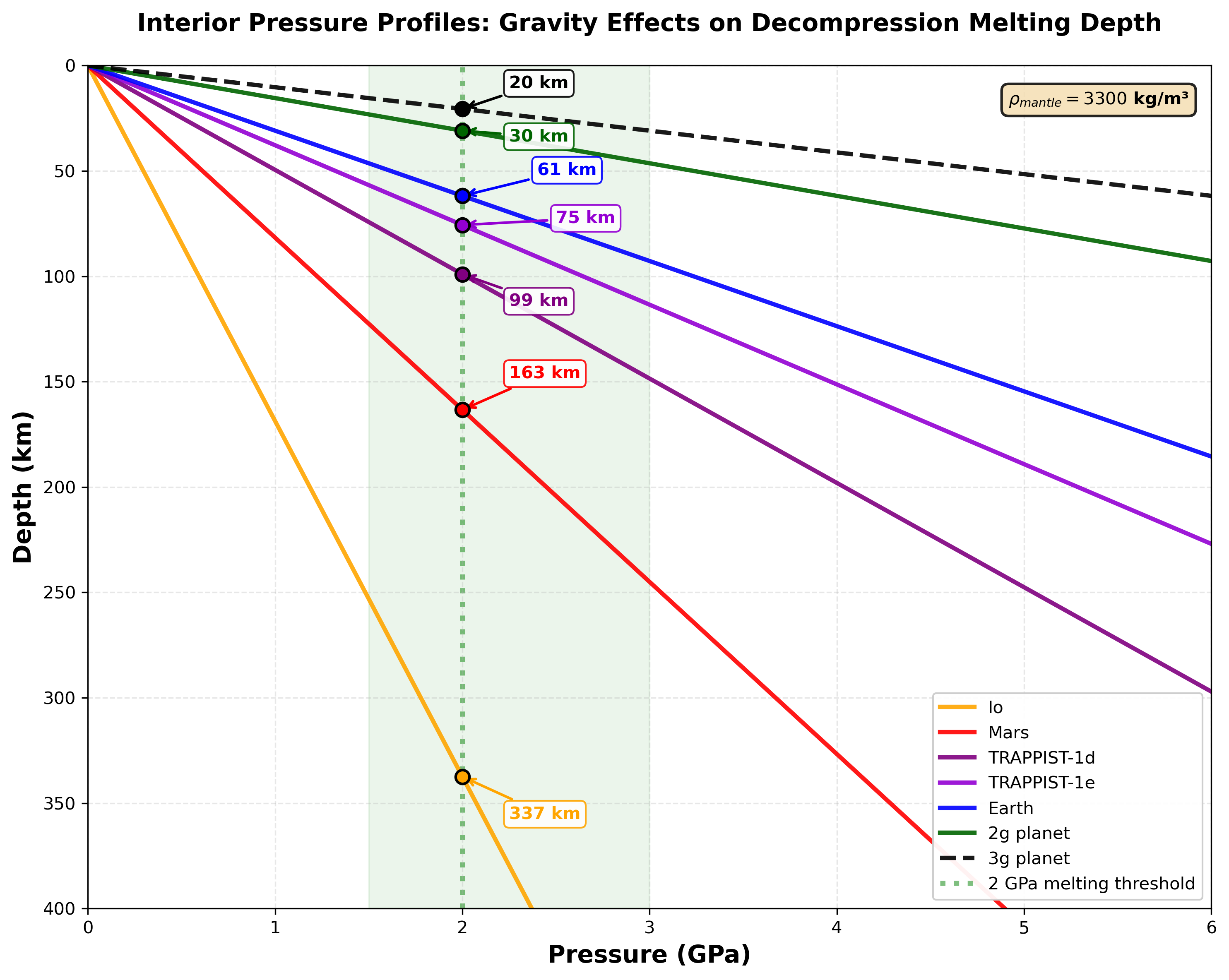}
    \caption{Interior pressure profiles as a function of depth for rocky bodies with varying surface gravities. Pressure increases linearly with depth according to $P(z) = \rho_{\text{mantle}} \times g \times z$, assuming a uniform mantle density of 3300 kg/m$^3$. The vertical green dashed line marks the 2 GPa pressure threshold for decompression melting onset in peridotite mantles. Colored circles indicate the depth at which each body reaches this melting threshold: higher gravity bodies (e.g., 3g planet at 20 km) reach melting conditions at shallower depths compared to lower gravity bodies (e.g., Io at 337 km). The green shaded region (1.5--3.0 GPa) represents the typical pressure range for sustained decompression melting. Hypothetical 2g and 3g planets are shown to illustrate the strong gravity dependence, with the 3g planet profile shown as a dashed line for reference.}
    \label{fig:pressure_profiles}
\end{figure}

Initial volcanic plume conditions used in this study that often control eventual plume heights are also explored as variables across a phase space, but they derive from magma that has been generated from the planetary interior.  Understanding this interior to plume connection is a complicated but fascinating underexplored topic in the context of exoplanet science, especially since the diversity in planetary properties of discovered exoplanets effectively offers a laboratory for this connection.  As one simple illustrative example, a potentially key controlling factor for some of the eruptions described here is the depth at which decompression melting occurs.  Decompression melting is likely the dominant mechanism for basaltic magma generation on the Earth over it's history through the present, and was a significant (and likely dominant) magma generation mechanism for most of the terrestrial bodies interior to the asteroid belt as well \citep{christiansen2023origin}. The depth of decompression melting has first-order implications for magma volatile content and eruption style that scale systematically with surface gravity (Figure~\ref{fig:pressure_profiles}). This relationship provides key insights into potential volcanic processes on diverse exoplanets. 

We use a toy model to show where rising mantle plumes would likely begin to experience significant decompression melting, approximated by a 2 GPa pressure threshold. Surface gravity strongly influences the depth at which this melting threshold is reached, with higher gravity planets achieving this pressure at much shallower depths. Our simplified model assumes a uniform mantle composition of peridotite with density $\rho_{mantle} = 3300$ kg/m³ for all bodies, and uses a simple hydrostatic equation $P = \rho_{mantle} \cdot g \cdot z$ to model interior pressure. While these simplifications allow for a clear comparative analysis, actual melting processes on exoplanets will be influenced by a complex interplay of factors including thermal evolution, compositional heterogeneities, and tectonic regimes. Future work incorporating more detailed planetary interior models, volatile solubility models, magma ascent rates as functions of gravity and tidal heating, and crustal thickness variations could refine predictions of which exoplanetary environments favor explosive versus effusive volcanism and efficient atmospheric volatile delivery.

Even such a simple model may be able to elucidate key principles. High-gravity super-Earths (2--3~$g_{\oplus}$) experience shallow melt generation (20--30~km depth), where lower pressures reduce volatile solubility in magmas and can yield volatile-poor melts compared to Earth. Additionally, higher gravitational acceleration opposes explosive fragmentation, requiring substantially higher overpressures to drive explosive eruptions. Together, these effects suggest that high-gravity rocky exoplanets may be dominated by effusive volcanic flows rather than explosive eruptions, with efficient atmospheric injection requiring extreme vent conditions. When combined with our finding that dense atmospheres demand high initial momentum to overcome entrainment penalties, volcanic atmospheric injection on high-gravity worlds faces a compounding challenge: shallow, volatile-depleted sources must generate sufficient overpressure to overcome both gravitational resistance and atmospheric drag.

Conversely, low-gravity bodies like the TRAPPIST-1 planets (0.6--0.8~$g_{\oplus}$) and smaller rocky worlds reach melt-initiation pressures ($\sim$2~GPa) at substantially greater depths ( Figure~\ref{fig:pressure_profiles}), where high-pressure conditions favor greater volatile incorporation into ascending magmas. This depth of initial melting has important implications for resulting magma properties: volatile abundance may be higher in magmas originating from greater depths, plume temperature is influenced by the extent of adiabatic decompression, and initial eruption velocities might be higher for magmas rising from greater depths due to increased potential energy.  However, this complicated by the fact the long ascent path presents competing effects: volatiles may be retained if ascent is rapid through established conduits (as may occur with tidal heating), enabling highly explosive eruptions that can efficiently inject material into thin atmospheres under low-gravity conditions. Alternatively, slow ascent through thick stagnant lids could result in extensive degassing en route, producing effusive buoyancy driven eruptions at the surface despite volatile-rich source regions. 

The interplay between melt generation depth, volatile content, ascent dynamics, and gravitational effects suggests that surface gravity fundamentally shapes not only plume dynamics (as demonstrated in our simulations) but also the pre-eruptive conditions that determine eruption style and volatile budgets. Rocky exoplanets span a factor of $\sim$20 in surface gravity (1.5--30~m\,s$^{-2}$), with corresponding variations in decompression melting depths spanning over an order of magnitude (Figure~\ref{fig:pressure_profiles}), implying that volcanic processes and their atmospheric impacts may vary dramatically across the population of potentially observable rocky worlds.  Most critically, even simple models of the connection between the interior and volcanic eruptions for exoplanets may be able to produce testable predictions for observations.  Given the large and diverse sample of putatively rocky exoplanets that may eventually be amenable to observation, they may be powerful means of understanding how volcanism operates across different parameters.

\subsection{Forward Modeling of Volcanism in Close-in Exoplanet Systems}
The recent detection of SO$_2$ in exoplanet atmospheres with JWST \citep[e.g., in WASP-39b and warm Neptunes;][]{Tsai2023,Banerjee_2024,Gressier_2024,Gressier2025} and the demonstration that transient sulfate aerosols from large eruptions could be detectable in transmission \citep{Misra2015} motivate exactly this type of forward modeling. Our results show that on TRAPPIST-1e–like worlds, explosive plumes driven by Io-level or higher tidal heating can plausibly inject sulfur-bearing gases and condensates directly into the limb-probed pressures, where they may produce time-variable absorption features or continuum slopes analogous to those predicted in earlier sulfate-aerosol studies \citep{Misra2015,Quick2020}. Conversely, parameter combinations that confine plumes to pressures $\gtrsim 1$~bar are unlikely to produce a strong direct signature in transmission, although they may still shape the global climate and surface environment.

\subsection{From tidal heating to spectra: a comparative-planetology framework}
Our plume experiments, though anchored to a single TRAPPIST-1e atmospheric column or analytic profiles, have broader implications for assessing volcanism on tidally heated terrestrial exoplanets. A growing body of work suggests that many close-in rocky and sub-Neptune planets experience substantial internal dissipation, with Io-like heat flows expected for a non-negligible fraction of systems \citep{Jackson2008,Henning2009,Barr2018,Quick2020,Barr2023}. In several cases, volcanic or sulfur-rich atmospheres have already been identified as plausible or even favored interpretations of current data, for example in the context of transient sulfate aerosols \citep{Misra2015}, sulfur-rich atmospheres on tidally heated planets \citep{Jackson2008,Misra2015}, and recent JWST inferences of SO$_2$ and other sulfur species in exoplanet atmospheres \citep[e.g.,][]{Tsai2023,Banerjee_2024,Gressier_2024,Gressier2025}. In parallel, interior and tidal-heating studies of the TRAPPIST-1 system indicate that several planets, including TRAPPIST-1e, f, g and h, may sustain long-lived volcanic or cryovolcanic activity \citep{Barr2018,Quick2020,Nicholls2025}.

Within this context, our simplified plume model provides a physically interpretable bridge between a given interior heat-flow scenario and its possible atmospheric signature. By injecting a continuum of idealized explosive plumes into a realistic TRAPPIST-1e background column, we map source parameters (vent pressure excess, exit velocity, volatile mass fraction) into diagnostic plume-top pressures and heights. In our fiducial 2~bar CO$_2$ column, representative plumes typically rise from the near-surface (order 1--2~bar) to altitudes of tens of kilometers, corresponding to pressures of order $10$--$10^{-2}$~bar depending on the case. The highest plumes in our suite overshoot to pressures of a few $\times 10^{-2}$~bar, while weaker or more heavily entraining plumes stall deeper, around $10^{-1}$--$1$~bar.

This makes the current model a flexible tool for comparative planetology. Even though we have so far used a single TRAPPIST-1e GCM column, the key controlling quantities---vent overpressure, exit Mach number, bulk volatile content and latent heating, Brunt--V\"ais\"al\"a frequency profile $N(z)$, and background wind shear---can be reinterpreted for a wide class of tidally heated exoplanets. Interior and orbital models already provide estimates of tidal heat fluxes and plausible tectonic/volcanic regimes for many systems \citep{Jackson2008,Henning2009,Barr2018,Quick2020,Nicholls2025}. Our framework therefore makes it possible to translate interior scenarios into testable atmospheric predictions: for a given planet mass, radius and incident flux, one can (i) prescribe a background $T$--$p$ column consistent with existing climate or radiative--convective models, (ii) choose plume source parameters compatible with the inferred internal heat budget, and (iii) compute whether the resulting plume heights intersect the pressures to which JWST and future observatories are most sensitive.

\subsection{Volcanic Plumes as a Strong Atmospheric Signature for Close-in Exoplanet Systems}
A central implication of our results is that explosive volcanism can perturb atmospheric columns at pressures far lower than those typically accessed by meteorological cloud decks, and therefore can imprint a disproportionately large signature on transmission and emission observables. Terrestrial cloud decks are largely constrained by moist convection and radiative cooling within the troposphere, whereas volcanic plumes can inject gases and particulates into the stratosphere/mesosphere and, in extreme cases, to pressures approaching the onset of the collisionless regime. Once material is deposited at sufficiently low pressure (high altitude), it resides at large geometric altitudes where the slant optical path sampled in transmission becomes enormous; even modest column masses of aerosols can therefore yield large changes in the effective tangent optical depth. In practice, this implies that volcanic plumes may influence observables not only through molecular features (e.g., \ce{H2O}, \ce{SO2}, \ce{CO2} depending on composition), but also through broadband opacity from aerosols (and potentially condensates) that can mute or reshape the spectrum in ways that are difficult to replicate with deeper tropospheric clouds.

These pressure ranges are directly relevant for observability. Transmission spectroscopy is primarily sensitive to atmospheric structure and composition over a relatively narrow range of pressures near the planetary limb, typically $10^{-3}$--$1$~bar for strongly irradiated giant planets, and somewhat deeper for high-mean-molecular-weight or cooler atmospheres where the scale height is smaller and clouds/hazes truncate the limb \citep{Burrows2014,Irwin2014,Molliere2017}. Emission spectra at secondary eclipse and in phase curves tend to probe photospheric levels closer to $\sim 0.1$--1~bar, depending on wavelength, opacity, and irradiation \citep{Burrows2014,Huitson2012,Coulombe2023}. In that sense, our TRAPPIST-1e plume experiments occupy a ``sweet spot'': a substantial fraction of the volcanic material is lofted into the 10--100~mbar region, where both transmission and emission spectra can be strongly modulated by additional absorbing gases and aerosols, but the plumes remain dynamically coupled to the lower atmosphere rather than escaping into the exosphere.

However, transmission spectroscopy integrates the full limb, not a single vertical column above the vent. This requires modeling the 3D redistribution of plume material by atmospheric circulation. For synchronously rotating terrestrial planets, the relevant observables depend on (i) where eruptions occur relative to the day--night pattern, (ii) how quickly the circulation exports material to the terminators, and (iii) how long high-altitude aerosols persist against sedimentation and microphysical growth. Because the transmission spectrum effectively averages azimuth around the terminator, a localized plume will produce a detectable signature only if it both reaches sufficiently low pressures and spreads over a substantial fraction of the limb before it is removed or diluted.

A crucial next step, which we reserve for a follow-up study, is to relax the 1-D, steady-plume approximation and embed these volcanic sources self-consistently into full 3-D general circulation models. This will allow us to track how jets and overturning circulations redistribute volcanic gasses horizontally and vertically, how microphysics converts them into aerosols, and how radiative feedback from ash and sulfate clouds modify both the background climate and subsequent plume evolution \citep[e.g.,][]{Misra2015,Quick2020}. Coupling our plume module to GCMs such as ExoCAM \citep{Wolf_2022} will make it possible to generate synthetic time-resolved spectra (transmission, eclipse, and phase curves) for tidally heated terrestrial planets, directly linking interior heating scenarios to observable atmospheric variability in the era of JWST and the forthcoming Habitable Worlds Observatory (HWO). In this sense, the single TRAPPIST-1e column used here should be viewed not as a limitation, but as a calibrated benchmark from which a broad family of tidally forced, volcanically active exoplanets can be explored within a unified physical framework.

\subsection{Opacity is microphysics-limited and composition-controlled}
Observability hinges not only on how high plume material rises, but on what optical properties it develops as it evolves. Aerosol loading, particle size, and vertical distribution are not free parameters in nature. They depend on plume composition (e.g., sulfur-bearing vs.\ water-rich vs.\ ash-dominated), local thermodynamics, and the ambient atmospheric state (temperature structure, background composition, and condensable reservoirs). The same injected mass can be nearly transparent (rapid growth and fallout) or highly opaque (sustained small-particle populations). This motivates coupling ExoCAM to an aerosol microphysics framework such as CARMA \citep{Gao_2021} to predict time-dependent particle size distributions and vertical profiles via nucleation/condensation, coagulation, sedimentation, and mixing. Such a coupling provides a physically grounded mapping from eruption composition to refractive indices and wavelength-dependent opacity, which is essential to interpret broadband flattening versus targeted molecular features.

\subsection{Context from recent JWST constraints in the TRAPPIST-1 system}
Recent JWST analyses of TRAPPIST-1 provide an important backdrop for selecting plausible background atmospheres and for assessing the degree to which stellar contamination can mask subtle signals.

For TRAPPIST-1e, \citet{Glidden_2025} analyze JWST/NIRSpec PRISM transmission spectra and report no strong evidence for or against an atmosphere. They weakly disfavor \ce{CO2}-rich atmospheres at pressures corresponding to Venus/Mars surface conditions and Venus cloud-top pressures, exclude \ce{H2}-rich atmospheres containing \ce{CO2} and \ce{CH4}, and find that higher-mean-molecular-weight, \ce{N2}-rich atmospheres with trace \ce{CO2} and \ce{CH4} remain permitted. Both an airless (bare rock) scenario and a high-$\mu$ \ce{N2}-rich scenario can fit the data, with residual features plausibly attributable to uncorrected stellar contamination and/or weak atmospheric signals.

Complementary constraints for TRAPPIST-1d come from \citet{Piaulet-Ghorayeb_2025}, who present a 0.6--5.2~$\mu$m NIRSpec/PRISM transmission spectrum from two transits. They find that the transmission spectrum is consistent with being flat at the $\sim$100--150~ppm level, showing no clear haze-like slope or molecular absorption. Their analysis rules out clear \ce{H2}-dominated atmospheres and rejects a range of higher-mean-molecular-weight analogs (e.g., Titan- and cloud-free Venus/Mars-like scenarios) at high confidence, implying that if an atmosphere exists it is either extremely thin and/or involves high-altitude aerosols.

These findings motivate two practical points for exovolcanism observability in the TRAPPIST-1 system. First, stellar contamination can be a limiting systematic at amplitudes comparable to (or larger than) expected volcanic spectral perturbations, so detectability may rely on multi-epoch strategies and contamination-informed retrievals. Second, if high-altitude aerosols are common (whether from climatological clouds/hazes or episodic volcanism), then time variability and wavelength-dependent opacity diagnostics become central discriminants.

\subsection{Observational outlook}
Even if TRAPPIST-1e ultimately proves to be airless or hosts an atmosphere outside the specific cases modeled here, our conclusions remain portable because the dominant control parameters are (i) the maximum altitude (minimum pressure) reached by plume material, (ii) the wavelength-dependent opacity of its gases and aerosols, and (iii) the 3D covering fraction and lifetime of high-altitude material at the limb. Background composition primarily modulates these through scale height, stability/cold traps, and chemistry/microphysics. TRAPPIST-1d/e are therefore best viewed as convenient, observationally accessible testbeds in which to stress-test plume injection physics and evaluate detectability in regimes that will generalize to other temperate rocky worlds.

\begin{figure}[H]
    \centering
    \includegraphics[width=0.85\textwidth]{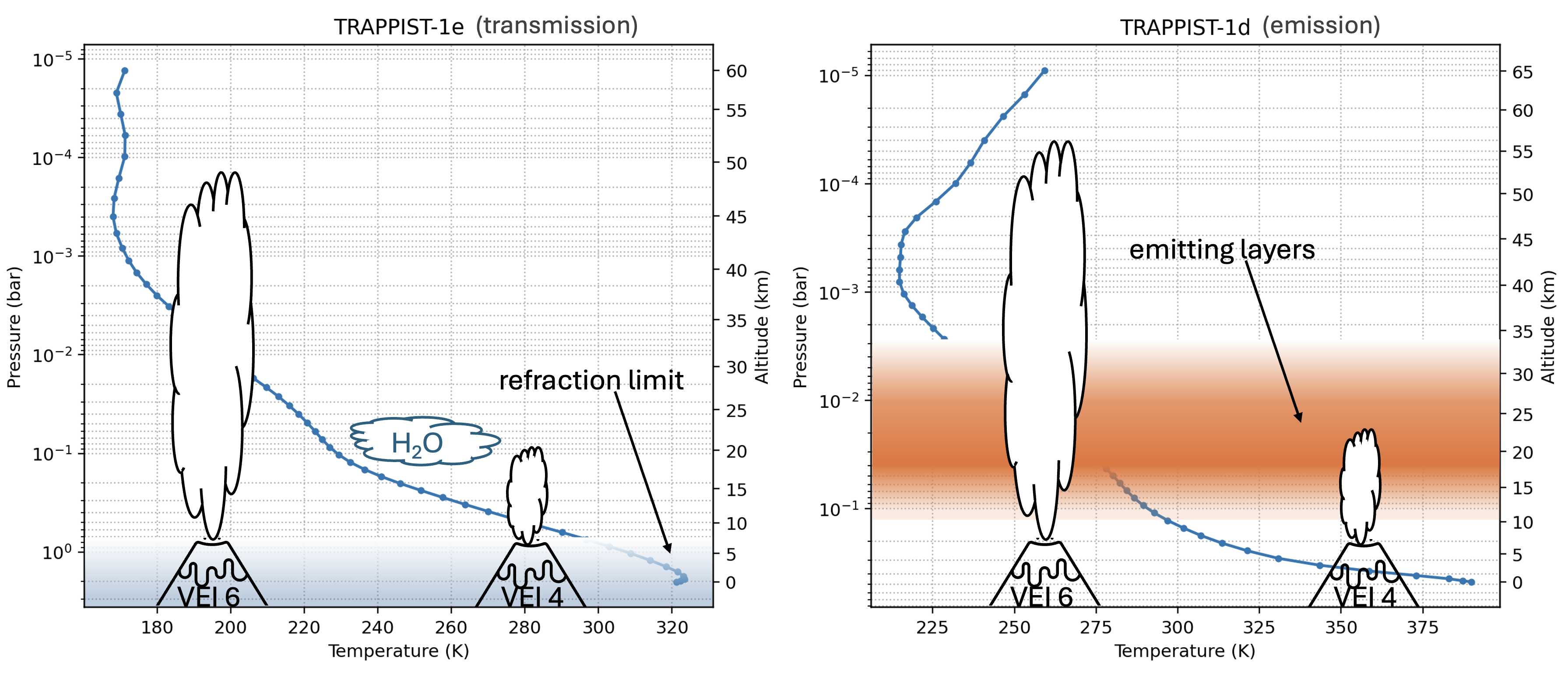}
    \caption{Schematic figure of exovolcanic plume detectability in transmission for TRAPPIST-1e (left) and emission for TRAPPIST-1d (right). The refraction limit at $\lambda = 1~\mu m$ has been computed using the PSG \citep{Villanueva_2022} ray-tracing calculation. The emission layers of TRAPPIST-1d have been estimated using the ExoCAM 0.5~bar CO$_2$ profile at the substellar point and the Eddington–Barbier approximation, i.e., that the emergent intensity at a given wavelength (we consider here a 13--18 $\mu m$ window) primarily originates from the atmospheric layers where the vertical optical depth to space is of order unity ($\tau_\lambda \approx 1$), often ($\sim 2/3$). Under hydrostatic balance and assuming the mass absorption coefficient varies slowly over the contribution region, ($\tau_\lambda(p)\simeq \kappa_\lambda p/g$), so the effective emitting pressure scales as ($p_{\rm emit}\sim g/\kappa_\lambda$); consequently, the high opacity in the CO$_2$ $15~\mu m$ band core shifts emission to mbar levels, whereas the lower-opacity wings probe deeper pressures (tens of mbar to $\sim0.1$ bar).
}
    \label{fig:sketches}
\end{figure}

Taken together, our results, schematically represented on Fig. \ref{fig:sketches}, imply that beyond VEI 4 volcanic aerosols injected above the typical refraction limit and cloud-forming region can be among the most observationally consequential outcomes of explosive volcanism on terrestrial exoplanets. Because transmission spectra are especially sensitive to high-altitude opacity, the most promising signatures may be time-variable continuum suppression/flattening and/or episodic molecular bands associated with plume volatiles---provided stellar contamination is controlled and the plume achieves sufficient limb coverage. For warm/hot rocky exoplanet like TRAPPIST-1d (right side of Fig. \ref{fig:sketches}), volcanic material injection from VEI 4 largely covers and expands above the emission layers, suggesting potential detectability in emission spectroscopy. Beyond those qualitative assessments, the next modeling pathway is therefore clear: combine 3D GCM transport (e.g., ExoCAM) with composition-driven aerosol microphysics (e.g., CARMA) and forward-model the full observing geometry to determine when a single eruption is detectable in practice versus when only sustained, frequent volcanism yields a statistically significant imprint. This will be the subject of a follow-up paper, currently in preparation.

\begin{nolinenumbers}

\begin{acknowledgments}
 P.S. and T.J.F. acknowledge that this work was supported by the National Aeronautics and Space Administration (NASA) under the Research Opportunities in Space and Earth Sciences (ROSES) program Habitable Worlds, grant 80NSSC23K1524. P.S. and T.J.F thank Eric T. Wolf and Joe Renaud for productive discussions that motivated this paper. P.S. and T.J.F. acknowledge support from NASA GSFC’s Sellers Exoplanet Environments Collaboration (SEEC). We acknowledge the use of ChatGSFC, an AI-assisted research tool, for its assistance in figure preparation and formatting of this manuscript. The authors were solely responsible for all scientific ideas, analysis, modeling, and interpretation of results presented in this work. P.S thanks Rose S. for inspiring and providing support on the effort required for the modeling.
\end{acknowledgments}

\end{nolinenumbers}

\bibliography{main}{}
\bibliographystyle{aasjournal}

\end{document}